\documentclass[a4paper,11pt]{article}
\pdfoutput=1 % if your are submitting a pdflatex (i.e. if you have
\usepackage{jcappub} % for details on the use of the package, please
\usepackage[T1]{fontenc} % if needed

\usepackage{comment}

\newcommand{\sH}{\mathcal{H}}
\newcommand{\p}{\prime}

\title{Effective field theory of magnetogenesis identify necessary and sufficient conditions}
\author[a,1]{Ashu Kushwaha,\note{Equal contribution to this work.}}
\author[a,2]{Abhishek Naskar,\note{Equal contribution to this work.}}
\author[b]{Debottam Nandi,}
\author[a,3]{and S. Shankaranarayanan\note{Manuscript correspondence should be addressed to.}}

\affiliation[a]{Department of Physics, Indian Institute of
  Technology Bombay, Mumbai 400076, India}
\affiliation[b]{Department of Physics \& Astrophysics, University of Delhi, Delhi 110007, India}
\emailAdd{ashu712@iitb.ac.in}
\emailAdd{30004198@iitb.ac.in}
\emailAdd{dnandi@physics.du.ac.in}
\emailAdd{shanki@iitb.ac.in}

\abstract{At astrophysical and cosmological scales, there is a detectable amount of magnetic field. There are several probable origins for this observed magnetic 
field, including the possibility of its origin in the early Universe. There are several models for primordial magnetogenesis, and if the inflationary background is taken into account, broken conformal invariance is required to generate a sufficient amount of magnetic field.  The breaking of conformal invariance is introduced either by new couplings 
between electromagnetic field and inflaton field or including higher derivative terms to the theory. 
As a step to unify these different approaches in the literature, we propose an Effective Field Theory (EFT) approach based on expansion about the Hubble parameter $(H)$ and its derivatives, where EFT parameters describe the magnetogenesis scenario in the early Universe, and different choices of parameters correspond to different models. 
We explicitly show that the generation of primordial magnetic fields requires two necessary conditions --- conformal invariance breaking and causal propagation. While broken conformal invariance is a common requirement for primordial magnetogenesis, for the first time, we show that causal propagation is also a necessary condition. We confirm this by considering a specific model of primordial magnetogenesis.}

\begin{document}
\maketitle
\flushbottom

\section{Introduction}

Various observations have confirmed the existence of magnetic fields in the Universe~\cite{1994-Kronberg-Rept.Prog.Phys.,2001-Grasso.etal-PhyRep,2010-Neronov.Vovk-Sci}. In galaxies and galaxy clusters, the typical magnetic field strength is found to be on the order of micro-Gauss with a coherence length of kpc to Mpc~\cite{1994-Kronberg-Rept.Prog.Phys.,2001-Grasso.etal-PhyRep,2002-Widrow-Rev.Mod.Phys.,2004-Giovannini-IJMPD,2013-Durrer.Neronov-Arxiv,2016-Subramanian-Arxiv}. Existing data on the magnetic fields in these regions cannot directly constrain the properties and origin of cosmic-scale magnetic fields. Therefore, it is unknown whether their origin is astrophysical or primordial. However, magnetic field measurements from Faraday rotation and Synchrotron radiation provide an upper bound for magnetic fields. In contrast, FERMI measurements of gamma-rays emitted by Blazars provide a lower bound of the order of $10^{-15}~{\rm G}$ in the intergalactic voids~\cite{2010-Neronov.Vovk-Sci}. 

According to the widely accepted paradigm, magnetic fields in these regions are produced by the dynamo amplification of the weak primordial magnetic field~\cite{2016-Subramanian-Arxiv}. There are various mechanisms for producing the primordial magnetic field, but in most of them, either the produced magnetic field is too weak to be amplified via dynamo or the coherence length is too short to be sustained due to Universe expansion~\cite{2016-Fabre.Shanki-APP}. Inflation provides a causal mechanism to generate the magnetic field over a large scale~\cite{1988-Turner.Widrow-PRD}. 
However, one of the pre-requisites for generating the primordial magnetic field during inflation is breaking the conformal invariance of the 4-D electromagnetic action. 

Several models have been proposed to break the conformal invariance of the action without breaking the gauge invariance. Broadly they can be classified into two categories --- coupling electromagnetic fields with other matter (scalar) fields and higher-derivative terms in the electromagnetic action leading to the non-minimal coupling of the electromagnetic field with curvature~\cite{1988-Turner.Widrow-PRD,1991-Ratra-Apj.Lett,1993-Dolgov-PRD,2020-Talebian.etal-arXiv,2020-Bamba.Odintsov.etal-JCAP,2021-Giovannini-JCAP,2014-Basak-Shanki-JCAP,2017-Debottam.Shankaranarayanan-JCAP,2019-Kushwaha.Shankaranarayanan-PRD,2020-Kushwaha.Shankaranarayanan-PRD}. Due to simplicity, the first class of models, especially, scalar field coupled models are extensively studied~\cite{1991-Ratra-Apj.Lett,2020-Talebian.etal-arXiv}. However, these models suffer from strong coupling and back-reaction problems that necessitate parameter tweaking. 
For the possible resolution of these issues, see Refs.~\cite{2009-Demozzi.etal-JCAP,2017-Sharma.etal-PRD,2021-Nandi-JCAP,2021-Tripathy.etal-arXiv}. 

The second class of models is more natural as higher derivative terms are expected to arise when quantum gravitational effects are taken into account~\cite{1981-DeWitt-PRL,1996-Padmanabhan-PRL}. Demanding that the theory be Lorentz invariant in flat space-time, the field action (in Fourier space) can only be a function of 
$k^2 (\equiv k_{\mu} k^{\mu})$~\cite{1983-Barth.Christensen-PRD,1990-Simon-PRD,2002-Hawking.Hertog-PRD}. Besides, the divergence
structure of quantum field theory is expected to vastly improve when
the quantum gravitational effects are taken into account~\cite{1983-Barth.Christensen-PRD}. For instance, higher-derivative electromagnetic theory by Podolsky-Schwed~\cite{1948-Podolsky.etal-RevModPhys} removes the divergence of the Coloumb potential. One problem with higher derivative theories is the appearance of negative energy states. Although they can be traded by negative norm states (or ghosts), they normally lead to non-unitary theories. However,  vector Galileons do not have ghosts~\cite{2017-Debottam.Shankaranarayanan-JCAP,2019-Kushwaha.Shankaranarayanan-PRD}.

Ideally, we require a fundamental theory of quantum gravity to obtain a generic magnetic field power spectrum generated in the early Universe. However, since we do not have such a consistent model of quantum gravity yet, we aim to obtain an \emph{effective field theory} (EFT) description of primordial magnetogenesis during inflation (based on expansion about the Hubble parameter $(H)$ and its derivatives). In this work, we obtain a generic magnetic field power spectrum from a low-energy effective field theory of magnetogenesis. While broken conformal invariance is a common requirement for primordial magnetogenesis, for the first time, we show that causal propagation is also a necessary condition.

While EFT of inflation has been systematically analyzed following Ref.~\cite{2007-Cheung.etal-JHEP}, there is no such systematic analysis for magnetogenesis. See, for instance, Refs.~\cite{2021-Giovannini-PLB,2021-Maity.etal-JCAP}. 
However, two key differences exist between the EFT of magnetogenesis and inflation.  
First, the EFT of inflation is a model-independent framework for studying scalar (and tensor) perturbations. In this setup, 
time-translation invariance needs to be broken as inflation ends at a finite time. In the case of EFT of magnetogenesis, the conformal invariance of the gauge fields also needs to be broken in the cosmological background besides breaking time diffeomorphism. 
Second, in EFT of inflation, one makes a specific gauge choice (unitary gauge) 
where the scalar-field (inflaton) fluctuations are zero, add gravitational operators to the Lagrangian that preserves spatial-diffeomorphism and breaks the time diffeomorphism. As the system breaks time diffeomorphism, one can write down a Lagrangian for the Goldstone Boson associated with the broken symmetry with the Stuckleberg trick. Now, this Goldstone Boson $\pi$ is related to the gauge-invariant quantity the curvature perturbation $\zeta$ as $\zeta = - H \pi$ with $H$ being the Hubble parameter. So by analyzing the dynamics of $\pi$, one can analyze the scalar mode of perturbation through the gauge-invariant quantity $\zeta$ produced during inflation. 
\iffalse
\textcolor{red}{Based on these, one choice of EFT Lagrangian of Inflation is~\cite{2007-Cheung.etal-JHEP}:}
\textcolor{red}{
\begin{multline}\label{EFT-inf}
	\mathcal{S}=\int d^4x \sqrt{-g}\left[\frac{1}{2}M_{\rm Pl}^2 R-\Lambda(t)-c(t)g^{00}+\right. \\
           \left.  \frac{1}{2}M_2(t)^4(g^{00}+1)^2-\frac{\bar{M}_1(t)^3}{2}(g^{00}+1)\delta K_{\mu}^{\mu}
           -\frac{\bar{M}_2(t)^2}{2}\delta K_{\mu}^{\mu2}-\frac{\bar{M}_3(t)^2}{2}\delta K_{\mu}^{\nu} K_{\nu}^{\mu}\right. \\
            \left. +  \frac{M_3(t)^4}{3!}(g^{00}+1)^3-\frac{\bar{M}_4(t)^3}{3!}(g^{00}+1)^2\delta K_{\mu}^{\mu}
            -\frac{\bar{M}_5(t)^2}{3!}(g^{00}+1)\delta K_{\mu}^{\mu 2}-+..\right]\, ,
 \end{multline}}
%
\textcolor{red}{where $\delta K_{\mu \nu}$ is extrinsic curvature and $M_i(t)$ are arbitrary functions of the cosmic time $t$.}
\fi
%
%
In the case of magnetogenesis, we do not have to make any specific gauge choice where the perturbed gauge fields vanish. This is because, unlike the scalar perturbations, 
the vector modes of the gravitational operators do not 
dynamically affect the gauge field. Hence, we do not need to construct any gauge-invariant variable out of the gauge field and the vector modes of perturbation, and for the EFT expansion, we need only to consider the gauge field $A_{\mu}$. In Appendix \eqref{sec:Decoupling}, we show this explicitly for general relativity.%; to our knowledge, this has not been shown earlier in the literature. 

In this work, we systematically write the EFT of magnetogenesis in the early Universe in terms of the Hubble parameter $H(t)$ and its derivatives. More specifically, we expand the Lagrangian in the powers of the cut-off scale $\Lambda$ and consistently analyze the conditions for generating a primordial magnetic field without any other assumptions. 
These terms containing $\Lambda$ break the conformal invariance of the gauge field and satisfy one of the key criteria. Our approach is different from the approaches in Ref.~\cite{2021-Maity.etal-JCAP, 2021-Giovannini-PLB}. In Ref.~\cite{2021-Giovannini-PLB} the effective action terms only consist of four derivatives associated with functions of background inflaton. This expansion leads to different susceptibilities for 
electric and magnetic fields, but the results are produced with a particular parametrization of these susceptibilities. In Ref.~\cite{2021-Maity.etal-JCAP}, the EFT Lagrangian is written in second-order with all possible contractions of electromagnetic tensor $F_{\mu\nu}$ with itself associated with time-dependent analytic functions $f_i(\eta)$. The functions $f_i(\eta)$ is chosen to be proportional to either
$\left({a(\eta)}/{a_f(\eta_f)}\right)^2$ or higher powers where $\eta_i$ and $\eta_f$ are (conformal) time at beginning and end of inflation. 
More specifically, in both these cases, the authors did not
include terms that can be proportional to $({H}/{\Lambda})$ where in our expansion scheme, it naturally arises. 

The EFT method we use to study the physics of magnetogenesis in model-insensitive. We explicitly show that the conformal invariance breaking is \emph{only} a necessary condition, not a sufficient condition.  We show that we can have large amplifications even for super-luminal fluctuations. To avoid EFTs with superluminal fluctuations, we need another physical condition --- the modes should be sub-luminal. As explained in Ref.~\cite{2006-Adams.etal-JHEP}, local quantum field theories contain a Lorentz-invariant concept of causality and satisfy the typical S-matrix axioms. This should also be satisfied by the expansion scalar functions (cf. Eq. \eqref{EFT:L}) describing effective field theory of magnetogenesis. By construction, these functions are arbitrary, one need to impose additional conditions for a well-defined relativistic field theory. For a recent discussion in the context of black holes, see Ref.~\cite{2022-Serra.etal-JHEP}.
%We then show that the speed of propagation of the perturbed gauge fields need not be less than $1$. 
Thus, there are two necessary conditions for the generation of primordial magnetic fields --- conformal invariance breaking and causal propagation. We confirm this by considering a specific model of primordial magnetogenesis. 

The rest of the article is organized as follows: 
In section \ref{sec:EFT} we write down the EFT action of magnetogenesis and compare it with models proposed in the literature.  
Section \ref{sec:EFT-Spectrum} obtains the magnetic power spectrum for the EFT in slow-roll inflation and identifies the loophole in the magnetogenesis model building.  
In section \ref{sec:vector_galileon_example}, we take a specific example and show that magnetic field amplification is possible with the cost of super-luminal propagation. 
We present our conclusions and future prospects in section \ref{sec:discussion}. The five Appendices (\ref{sec:Decoupling} - \ref{app:VG-slow-roll}) contain details of the calculations.

In this work, we use $(+,-,-,-)$ metric signature and natural units where $\hbar = c = 1/(4\pi\epsilon_0) = 1$, with reduced Planck mass $M_{\rm Pl}^2 = 1/(8 \pi G) = 2.4 \times 10^{18}~{\rm GeV}$. The various physical quantities with the over-line refers to the values evaluated for the homogeneous and isotropic FRW background. A dot denotes a derivative with respect to the cosmic time ($t$), a prime stands for a derivative with respect to conformal time ($\eta$), and $,i$ denotes a derivative w.r.t spatial coordinates.

\section{EFT action of magnetogenesis}
\label{sec:EFT}

Like any effective field theory, the EFT of magnetogenesis includes two components: Symmetries and degrees of freedom~\cite{2007-Burgess-ARNPS,2020-Penco-Arxiv}. The vector perturbations 
do not influence the dynamical evolution of the gauge field. See Appendix \ref{sec:Decoupling} for details. Hence, the vector perturbations and gauge fields are decoupled, and the only relevant degree of freedom is the gauge field. Having identified the relevant degrees of freedom, our next step is to write down the action. 

In principle, effective action can have infinite terms. Therefore, we need to identify (broken) symmetries that describe the physics to expand the action about the expansion parameter.  Because of the conformal invariance of the standard electromagnetic action 
   \begin{align}
   \label{eq:SEM}
        S_{\rm EM} = -\frac{1}{4} \int d^4x \sqrt{-g} F_{\mu\nu} F^{\mu\nu} 
    \end{align}
it is impossible to produce a detectable amount of magnetic field from this setup\footnote{Note that $F_{\mu\nu} \tilde{F}^{\mu\nu}$ is a total derivative term in FRW background and does not contribute to the dynamics.}. Splitting the gauge-field in the background and perturbations \eqref{eq:VectPert}, we have:
\begin{equation}
    A_{\mu} = \bar{A}_{\mu} + \delta A_{\mu} = \delta A_{\mu}.
\end{equation}
Due to the background symmetry, we have considered the background field $\bar{A}_{\mu} =0$ and if we use the well-known choice of Coulomb gauge  (see Appendix \ref{sec:Decoupling} for details), 
\begin{equation}
\delta A_0 = 0, \partial_i \delta A^{i}=0 \, ,
\label{def:CoulombG}
\end{equation}
the Lagrangian for the fluctuation field $\delta A_{\mu}$ from \eqref{eq:SEM} can be written as,
\begin{align}
        S_{\rm EM} = \int d^4x \left[  (\delta A_i^{\prime})^2 - (\partial_i \delta A_j)^2 \right]
    \end{align}

This Lagrangian for the fluctuation is also conformally invariant and thus, the generation of primordial magnetic fields in the early Universe requires conformal invariance breaking of the electromagnetic action~\cite{1988-Turner.Widrow-PRD,1991-Ratra-Apj.Lett,1993-Dolgov-PRD}. In other words, the terms in the effective action must break the conformal invariance. Here, we demand that the effective action satisfies the following symmetries: 
\begin{enumerate}
\item \textbf{Local Lorentz invariance}: In the Minkowski limit, we demand that the effective action is Lorentz invariant. In other words, in the limit, $a(\eta) \to {\rm constant} (H(t) \to 0)$, the EFT action of magnetogenesis reduces to the standard electromagnetic action.
%second integral in the RHS of Eq.~\eqref{vec-L}.

\item \textbf{Gauge invariance}: As mentioned in detail in Ref.~\cite{2001-Jackson.Okun-RMP}, an implicit assumption behind the formalism of the gauge invariance is that the field equations must have unique solutions. We impose the validity of this condition in the early Universe. Thus, the EFT action \emph{do not} contain terms proportional to $A_{\mu} A^{\mu}$. Note that from now on, we will use $A_i$ instead of $\delta A_i$ to denote the gauge field fluctuation.
\end{enumerate}

While the above symmetry requirements will restrict the form of the EFT action of magnetogenesis, we need to identify the expansion parameters of the action. Although the two expansion parameters are well-defined in the time-independent system, it is not straightforward for time-dependent systems like in the early Universe~\cite{2017-Burgess-arXiv}.
In general, it is not possible to construct an EFT without energy conservation, as EFTs divide states based on energy. However, if the time-evolution of the degrees of freedom $(A_{i})$ is adiabatic --- they vary sufficiently slowly compared to the UV scales of interest $\Lambda$ --- we can treat the Hamiltonian obtained from the EFT action to be approximately conserved Hamiltonian with an approximate time-dependent low/high energy split~\cite{2015-Donoghue.Holstein-JPG,2017-Burgess-arXiv}. 

In the case of slow-roll inflation, the energy scale during inflation $H$ is approximately constant. Let us define energy scale $\Lambda$ that represents a cutoff scale for which the effects of high-scale physics are described by non-renormalizable operators, which can be thought of as originating from integrating out all particles with mass $m > \Lambda$~\cite{2003-Burgess-LivRev,2020-Penco-Arxiv}. Thus, at these energies, the scale dependence between the characteristic energy scale $H$ and the cutoff scale $\Lambda$ 
is given by the expansion parameter $H/\Lambda$. Like in general EFT, we fix the level of precision and sort by $H/\Lambda$ at a given order all terms that contribute to the action.
Note that we cannot use standard perturbation
theory to quantize, and the problem of non-renormalizability becomes an actual problem at energies close to $\Lambda$. This work assumes $\Lambda$ to be at least \emph{one order} higher than $H$ and less than $M_{\rm Pl}$.

Thus, the second-order EFT action of magnetogenesis, based on expansion about the Hubble parameter $(H)$ and its derivatives, is 
\begin{equation}
    \mathcal{S}_{\rm EFT} = \int d^4 x \left[f_1(H,a,\Lambda) ({A}_i^{\prime})^2 - f_2(H,a,\Lambda) 
    (\partial^j A_i  \, 
    \partial_j A_i )\right]
    \label{EFT:L}
\end{equation}
where, the \emph{expansion scalar functions} --- $f_1(H,a,\Lambda)$ and $f_2(H,a,\Lambda)$ --- depend of Hubble parameter $H$, scale factor $a(\eta)$, and cutoff scale $(\Lambda)$. This is a crucial expression regarding which we want to discuss the following: 
First, as explained in Ref.~\cite{2006-Adams.etal-JHEP}, local quantum field theories contain a Lorentz-invariant concept of causality and satisfy the typical S-matrix axioms. This should also be satisfied by the functions $f_1$ and $f_2$ describing effective field theory of magnetogenesis. By construction, these functions are arbitrary, one need to impose additional conditions for a well-defined relativistic field theory. For a recent discussion in the context of black holes, see Ref.~\cite{2022-Serra.etal-JHEP}.
Second, we have chosen the Coulomb gauge condition for the EM fields and hence resulting in only two physical degrees of freedom\footnote{The Coulomb gauge allows us to evaluate the observables like magnetic and electric power-spectrum efficiently.}. In principle, we can have separate expansion scalar functions in front of $A_1'^2, A_2'^2, A_3'^2$ but this will lead to different propagation speeds for the three components. (See Appendix \ref{general-propagation} for details). In this work, we focus on action \eqref{EFT:L}. 

Also, note that one can add parity breaking term  {$\epsilon^{ijk} A_i^{\prime} \partial_j  A_k$} in the EFT action (\ref{EFT:L}) with additional scalar expansion function $f_3$. In the limit of $\Lambda \to \infty $ and $a(\eta) \to {\rm constant}$ (or $H(t) \to 0$), $f_3 \to~\rm{constant}$. Since the parity symmetry only determines the nature of primordial magnetic field, we will not consider the parity breaking term in the action.
Third, in the limit of $\Lambda \to \infty$,
$f_1 = f_2 \simeq {\rm constant}$. In other words, the above expression will reduce to Eq. \eqref{eq:SEM}.
%second integral in the RHS of Eq.~(\ref{vec-L}). 
Also, in the limit of $a(\eta) \to {\rm constant}$ or $H(t) \to 0$), $f_1 = f_2 \simeq {\rm constant}$ and reduce to Eq. \eqref{eq:SEM}.
%second integral in the RHS of Eq.~(\ref{vec-L}).
%
Fourth, $f_1(H,a,\Lambda)$ and $f_2(H,a,\Lambda)$ capture all possible interactions of electromagnetic field that leads to breaking of conformal invariance.

Fifth, since $A_{\mu}$ has mass dimension $1$, both $f_1(H,a,\Lambda)$ and $f_2(H,a,\Lambda)$ have mass dimension zero~\cite{1992-Polchinski-arXiv}. 
Formally, the effective action can be written as~\cite{1992-Polchinski-arXiv}:
\begin{equation*}
    S \sim \int d^4 x ~ \mathcal{O}_{p,q} \sim \left(\frac{E}{\Lambda}\right)^{p +q - 4},
\end{equation*}
where, the operator $\mathcal{O}$ is made up of $p$ fields and $q$ derivatives. Since, all the observable quantities in cosmology are related to the Hubble parameter $H(t)$ and it determines the energy scale of the epoch. To compare the inflationary scale with EFT scale, a broad class of magnetogenesis models can be reproduced from the EFT that is a sum of series in $H/\Lambda$, time-derivatives of $H$ --- ($H'/\Lambda^2$), ($H''/\Lambda^3$), $\cdots$ --- and their products, for instance, $H H'/\Lambda^3$, $H H''/\Lambda^4$, $\cdots$. Since, $p$ and $q$ are integers, the effective field theory can be expanded only as a series in terms of the Hubble parameter ($\sH$) in conformal 
time\footnote{Higher derivatives of $\sH$ are also present, however, not shown in the expansion.}:
{\small
\begin{equation}
\label{eft:A}
\begin{split}
\!\!\!\!\!\!\! f_1(H, a, \Lambda) & = 
\sum_{n=0}^{\infty} s_n \frac{1}{a^n} \left(\frac{\sH}{\Lambda}\right)^n
    + \sum_{m=1}^{\infty} b_m \frac{1}{a^{2m}}
    \left(\frac{\sH^{\prime}}{\Lambda^2}\right)^m + \sum_{m,n=1}^{\infty} v_{n,m} \frac{1}{a^{m+2n}}
    \left(\frac{\sH}{\Lambda}\right)^m \left(\frac{\sH^{\p}}{\Lambda^2}\right)^n 
\\
\!\!\!\!\!\!\!  f_2(H, a, \Lambda) &= 
    \sum_{n=0}^{\infty} d_n \frac{1}{a^n} \left(\frac{\sH}{\Lambda}\right)^n
    + \sum_{m=1}^{\infty} e_m \frac{1}{a^{2m}}
    \left(\frac{\sH^{\prime}}{\Lambda^2}\right)^m + \sum_{m,n=1}^{\infty} w_{n,m} \frac{1}{a^{m+2n}} \left(\frac{\sH}{\Lambda}\right)^m \left(\frac{\sH^{\p}}{\Lambda^2}\right)^n 
\end{split}
\end{equation}
}
where $s_n, b_m, d_n, e_n, v_{n,m}, w_{n,m}$, are the unknown real (postive or negative) parameters and can be fixed for a particular magnetogenesis model. $s_0$ and
$d_0$ correspond to the values in standard electrodynamics satisfying the local Lorenz invariance. Note that $H' = (\mathcal{H^{\p}} - \mathcal{H}^2)/a$, hence the expansion in either of the two variables are equivalent. In the above expression, we have not included series in higher derivatives of $\sH$ like $(\sH''/\Lambda^3)$. In principle, these terms should also be included in the EFT and appear in higher-order gravitational coupling. As mentioned above, since the coefficients $s_n, b_m, d_n, e_n, v_{n,m}, w_{n,m}$ are unknown, one need to impose additional conditions, like causality, for a well-defined relativistic field theory~\cite{2006-Adams.etal-JHEP,2022-Serra.etal-JHEP}.

Sixth, in the literature, 
the odd powers of ${\sH}/{\Lambda}$ are not 
included and the first-order correction is taken to be ${\sH^2}/{\Lambda^2}$ or ${\sH^{\p}}/{\Lambda^2}$. However, we have included odd powers in
the expansion parameter to keep the analysis general. As mentioned above, the above expansion is valid only for $H < \Lambda$.

Lastly, Table (I) identifies the early Universe magnetogenesis models and the corresponding EFT parameters. The list is not exhaustive but gives a good representation of the various magnetogenesis models discussed in the context of inflation. Thus, we see that the EFT action \eqref{EFT:L} can reproduce most of the known magnetogenesis models. For most models, it is sufficient to consider up to the second order in the expansion parameter. Appendix \ref{app:coup-details} 
contains detailed calculations that provide a one-to-one mapping between the magnetogenesis model 
and EFT parameters.

\begin{table}[!htb]
\label{table:comparison}
\begin{tabular}{|c|c|}
\hline
{\bf Magnetogenesis models} & {\bf Non-zero EFT parameters} \\ \hline
Class of Ratra Model: $f(\phi) F_{\mu\nu}F^{\mu\nu}$  \cite{1991-Ratra-Apj.Lett} & 
$s_n, d_n$ (depending on $f(\phi)$) \\ \hline
Higgs Starobinsky Inflation  \cite{2022-Durrer.etal-arXiv} & 
$s_n, d_n$ \\ \hline
Vector Galileon Model~\cite{2017-Debottam.Shankaranarayanan-JCAP}        &      $s_2, e_1$          \\ \hline
Gravitational Coupling: $RF_{\mu\nu}F^{\mu\nu}$ \cite{1988-Turner.Widrow-PRD}&  $s_2, b_1, d_2, e_1$  %$= -3$
\\ \hline
Gravitational Coupling: $R_{\mu \nu}F^{\mu\alpha}F^{\nu}_{\alpha}$ \cite{1988-Turner.Widrow-PRD} & 
$s_2, b_1, d_2, e_1$
%$s_2=-\frac{1}{2}, b_1 =-1$ 
%and $d_2 =-1, e_1=-\frac{1}{2}$
\\ \hline 
Gravitational Coupling: $R_{\mu\nu\alpha\beta} F^{\mu\nu}F^{\alpha\beta}$
\cite{1988-Turner.Widrow-PRD} & 
$b_1, d_1$
%$b_1=d_1=-1$
\\ \hline 
Higher order Gravitational Coupling:  &  
$s_6, b_3, v_{4,1}, v_{2,2}$ \\
%$s_6 = -6, b_3 = -6, v_{4,1} = -18, v_{2,2} = -18$; \\
 $R^3 F_{\mu\nu}F^{\mu\nu}$ \cite{2022-Bertolami.etal-arXiv} & 
$d_6, e_3, w_{4,1}, w_{2,2}$ 
% $d_6 = -6, e_3 = -6, w_{4,1} = -18, w_{2,2} = -18$
\\ \hline 
\end{tabular}
\caption{One-to-one mapping between the magnetogenesis model and the EFT parameters. See Appendix \ref{app:coup-details} for details.}
\end{table}
Before we proceed with the rest of the analysis, we compare the above EFT action \eqref{EFT:L} with the ones recently proposed in the literature~\cite{2021-Giovannini-PLB,2021-Maity.etal-JCAP}. In Ref.~\cite{2021-Giovannini-PLB} the effective action terms only consist of four derivatives associated with functions of background inflaton. This expansion leads to different susceptibilities for 
electric and magnetic fields, but the results are produced with a particular parametrization of these susceptibilities. In Ref.~\cite{2021-Maity.etal-JCAP}, the EFT Lagrangian is written in second-order with all possible contractions of electromagnetic tensor $F_{\mu\nu}$ with itself associated with time-dependent analytic functions $f_i(\eta)$. The functions $f_i(\eta)$ are chosen to be proportional to either
$\left({a(\eta)}/{a_f(\eta_f)}\right)^2$ or higher-order where $\eta_i$ and $\eta_f$ are time at beginning and end of inflation. 
More specifically, in both these cases, the authors did not
include terms that can be proportional to $({H}/{\Lambda})$ where in our expansion scheme, it naturally arises. 
%Also, these authors do not identify the true degrees of freedom for magnetogenesis.

\section{Generic magnetic field power-spectrum from EFT action}
\label{sec:EFT-Spectrum}

In the previous section, we constructed EFT action \eqref{EFT:L} of magnetogenesis based on symmetries and degrees of freedom. We also constructed a form of the
expansion scalar functions ($f_1(H, a, \Lambda)$ and $f_2(H, a, \Lambda)$) and showed that the EFT parameters in this generic form indeed correspond to the various magnetogenesis models. To make the computation of the power-spectrum tractable and to highlight the importance of speed of perturbations, we truncate the series \eqref{eft:A} up to second order. However, the truncation of the series to compute the power spectrum has no bearing on the EFT expansion \eqref{eft:A}.
%\textcolor{red}{We also noted that for most models, it is sufficient to consider up to the second-order in the expansion parameter.}
%In this section, we derive the equation of motion (EOM) of the gauge field by truncating the expansion scalar functions to $\Lambda^{-2}$. We then obtain the power spectrum in the slow-roll inflation scenario. 

\subsection{Equation of motion from EFT action}

Truncating the expansion scalar functions ($f_1(H, a, \Lambda)$ and $f_2(H, a, \Lambda)$) in Eq.~\eqref{eft:A} to $\Lambda^{-2}$ order, we have,
\begin{equation}
\begin{split}
&   f_1(H, a, \Lambda) \simeq 
s_0 + \frac{s_1}{a(\eta)} \left(\frac{\sH}{\Lambda}\right) +  
\frac{s_2}{a^2(\eta)}\left(\frac{\sH}{\Lambda}\right)^2
+ \frac{b_1}{a^2(\eta)} \left(\frac{\sH^{\prime}}{\Lambda^2}\right),  \\
&  f_2(H, a, \Lambda) \simeq d_0 
+ \frac{d_1}{a(\eta)}\left(\frac{\sH}{\Lambda}\right) 
+ \frac{d_2}{a^2(\eta)} \left(\frac{\sH}{\Lambda}\right)^2
+  \frac{e_1}{a^2(\eta)} \left( \frac{\sH^{\prime}}{\Lambda^2}\right).
\end{split}\label{eft:truncated}
\end{equation}
As mentioned earlier, in the limit of $\Lambda \to \infty$, the expansion should reduce to Eq. \eqref{eq:SEM}.
%coefficients in the second integral in the RHS of Eq. \eqref{vec-L}. 
Hence, we have $s_0 = d_0 = 1/2$. (See Appendix \eqref{app:rhoEFT} for more details.)

To obtain the equation of motion corresponding to the action \eqref{EFT:L}, we first need to rewrite the action in canonical form. To do that, we define  
$A_i = \mathcal{A}_i/Z$ in the effective action \eqref{EFT:L} and we have: 
\begin{equation}
\label{EFT:L2}
 \mathcal{S}_{\rm EFT} = \int d^4 x \left[ (\mathcal{A}_i^{\prime})^2 + \frac{Z''}{Z}
    \mathcal{A}_i^2 - \frac{f_2}{f_1}(\partial_j \mathcal{A}_i)^2 \right].
\end{equation}
where $Z = f_1^{1/2}$. Using Eq. \eqref{eft:truncated}, we have: 
\begin{multline}
\frac{Z^{\p \p}}{Z} = 
\frac{s_1}{a(\eta)}
\left(\frac{\sH}{\Lambda}\right) 
\left[\frac{\sH^2}{2} - 3  \sH^{\prime} 
+ \frac{1}{2} \frac{\sH^{\p \p}}{\sH} \right] 
%%%
+ \frac{1}{a^2(\eta)}
\left(\frac{\sH}{\Lambda}\right)^2 
\left[ \left(\frac{s_1^2}{4}+ 2 s_2 \right) \sH^2  \right. \\ 
  \left. 
+ \left(s_1^2-5s_2+2b_1\right) \sH^{\p} 
+ \left( \frac{s_1^2}{4}+s_2-b_1 \right) 
\left(\frac{\sH^\p}{\sH}\right)^2 
- \left( \frac{s_1^2}{2}+s_2+2 b_1 \right) 
\frac{\sH^{\p\p}}{\sH} 
+ b_1 \frac{\sH^{\p\p\p}}{\sH^2}
\right] 
\end{multline}
Note that we have only kept terms up to $1/\Lambda^2$ and ignored higher-order $\Lambda$ contributions. Like in non-canonical scalar fields, $f_2/f_1$ in the effective action \eqref{EFT:L2} can be identified as the adiabatic sound speed:
\begin{eqnarray}
\label{def:CA}
c_A^2 &=& 
\frac{1+ \frac{d_1}{a(\eta)} \left(\frac{\sH}{\Lambda}\right) 
+ \frac{d_2}{a^2(\eta)} \left(\frac{\sH}{\Lambda}\right)^2
+ \frac{e_1}{a^2(\eta)} \left(\frac{\sH^{\prime}}{\Lambda^2}\right)}
{1+ \frac{s_1}{a(\eta)} 
\left(\frac{\sH}{\Lambda}\right) 
+ \frac{s_2}{a^2(\eta)} 
\left(\frac{\sH}{\Lambda}\right)^2 
+ \frac{b_1}{a^2(\eta)} \left(\frac{\sH^{\prime}}{\Lambda^2}\right)} \\
\label{def:CA2}
&\simeq& 1 + \frac{d_1-s_1}{a(\eta)} \left(\frac{\mathcal{H}}{\Lambda} \right)
+ \frac{1}{a^2} \left(\frac{\mathcal{H}}{\Lambda}\right)^2 \left[ 
s_1^2-s_2-s_1 d_1+d_2 + (e_1-b_1) (1 - \epsilon_1)
\right]
\end{eqnarray}
where in arriving at the above expression, we have assumed that $H/\Lambda$ is small and higher-order 
terms are negligible. This is the second key result of this work regarding which we want to stress the following points: First, as mentioned above, since the coefficients $s_1, s_2, d_1, d_2, b_1, e_1$ are unknown, one need to impose additional conditions, like causality, for a well-defined relativistic field theory~\cite{2006-Adams.etal-JHEP,2022-Serra.etal-JHEP}. More specifically, considering both $d_1$ and $s_1$ to be positive, the above expression implies that $c_A^2 > 1$ if $d_1 > s_1$, irrespective of the value of $H/\Lambda$. Thus, such models violate standard causality condition~\cite{2007-Ellis.etal-GRG}. Earlier,  effective field theories have been rejected based on super-luminal fluctuations as such
propagation generally leads to a global breakdown of causality~\cite{2006-Adams.etal-JHEP}. Second, models with $s_1 <0$ and $d_1 > 0$, will always lead to superluminal modes.
Third, when $d_1 = s_1$ and assuming all EFT parameters are positive, the causality condition implies $s_2 > d_2 + (e_1-b_1) (1 - \epsilon_1)$  during the entire inflationary epoch [$\epsilon_1$ is the first-order slow-roll parameter defined in Eq. \eqref{def:Slowroll}]. Lastly, in general, $c_A$ is a function of time. Since $H/\Lambda$ is small, one can assume that $c_A$ has a weak time dependence. Note that, by construction, the EFT action \eqref{EFT:L} is locally Lorenz invariant, and the causality condition imposes restrictions on the EFT parameters.

Fourth, in Appendix \eqref{app:rhoEFT} we have obtained the energy density \eqref{eq:EFT-rho} corresponding to the EFT action \eqref{EFT:L}. We infer the following from the energy-density \eqref{eq:EFT-rho}: $\rho_{\rm mixing}$ decays faster than $\rho_{\rm E}$ and $\rho_{\rm B}$. Hence, we can ignore $\rho_{\rm mixing}$ contribution in evaluating the energy density of the EFT. Take the extreme scenario where $\rho_{\rm E}$ and $\rho_{\rm B}$ contribute equally, imposing the condition that the energy density is always positive provides a condition that $s_1 > d_1$ (assuming $s_1, d_1$ are positive). This is consistent with the causality condition we obtained earlier. We show that all these features are satisfied for the specific Galileon vector model in Sec. \eqref{sec:vector_galileon_example}.

Lastly, the equation of motion corresponding to the action \eqref{EFT:L2}, in the Fourier domain $(k)$, is 
\begin{equation}
\label{eq:GFieldEOM}
\mathcal{A}_k ^{''} + \left[c_A^2 k^2 - \frac{Z^{''}}{Z} \right] \mathcal{A}_k = 0 \, ,
\end{equation}
where, $k = |{\bf k}|$ and ${\bf k}$ is the comoving wave vector. For brevity, we have defined $\mathcal{A}_k = \mathcal{A}_i^{(k)}$. In the rest of this section, we now compute the magnetic field power spectrum during inflation. 

\subsection{Generic magnetic power spectrum during inflation}

In this section, we quantize the effective gauge field given by the action \eqref{EFT:L2}, and obtain the general expression for the primordial magnetic field  (PMF) power spectra.  On quantization, the gauge field $\mathcal{A}$ can be expressed 
as follows:
\begin{eqnarray}\label{eq:fourierd}
\mathcal{\hat A}_i (\eta, {\bf x}) =  \int \frac{d^3 {\bf
     k}}{(2\pi)^{3/2}} \sum_{\lambda=1}^{2} \epsilon_{\lambda i}({\bf
   k}) \Big[\hat{b}_{\bf k}^\lambda \mathcal{A}_{\bf k} e^{i{\bf k.x}} +
   \hat{b}_{\bf k}^{\lambda\dagger} \mathcal{A}_{\bf k}^*(\eta) e^{- i{\bf k.x}}\Big]
 \, ,
 \end{eqnarray}
where $\lambda$ corresponds to two orthonormal transverse
polarizations, $\epsilon_{\lambda i}$ are the polarization
vectors and the creation (${\hat a}_{\bf k}$) and the annihilation (${\hat a}_{\bf k}^{\dag}$) operators obey the usual commutation relations. Like scalar and tensor perturbations~\cite{1992-Mukhanov.etal-Phy.Rep.},  the power spectrum as well as the statistical properties of the gauge field is characterized by the Wightman function of the gauge field. The power spectrum (the two-point correlation in Fourier space) is~\cite{2013-Durrer.Neronov-Arxiv}: 
\begin{equation}
\left\langle B_{i}^{*}(\mathbf{k}) B_{j}\left(\mathbf{k}^{\prime}\right)\right\rangle=(2 \pi)^{3} \delta^{3}\left(\mathbf{k}-\mathbf{k}^{\prime}\right) P_{i j} \mathcal{P}_{B}(k) \, ,
\end{equation}
where $P_{i j}$ is a projector onto the transverse plane and is given by
\begin{equation}
P_{i j}=\delta_{i j}-\frac{\mathbf{k}_{i} \mathbf{k}_{j}}{k^{2}},~~ P_{i j} P_{j k}=P_{i k}, ~~ P_{i j} \mathbf{k}^{j}=0 \, ,
\end{equation}
and $\mathcal{P}_{B}(|\mathbf{k}|)$ is the gauge field power spectrum. Since $B$ is statistically homogeneous and isotropic, the correlation depends only on the distance $|\mathbf{x}-\mathbf{y}|$.  
Using the decomposition (\ref{eq:fourierd}), the
PMF spectrum per logarithmic interval can then be written in terms of the modes $\mathcal{A}_{k}$ as
\begin{equation}
{\cal P}_{B}(k) 
= \frac{k^5}{2 \pi^2 a^4}\left\vert
\frac{\mathcal{A}_k}{Z}\right\vert^2
\label{eq:pMS}
\end{equation}
and the expression on the right-hand side is to be evaluated when 
the physical wavelength $(k/a)^{-1}$ of the mode corresponding to the comoving wavenumber ${\bf k}$ equals the \emph{effective sound horizon} of the electromagnetic fluctuations $c_A\,H^{-1}$~\cite{1999-Garriga.Mukhanov-PLB}.  The above condition translates to $(-k \eta) c_A = 1$ corresponding electromagnetic fluctuations exiting the sound horizon during inflation. This sound horizon is not the same sound horizon of the scalar perturbations during inflation~\cite{1999-Garriga.Mukhanov-PLB,2011-Hu-PRD}.
The normalization constant takes into account both modes of polarization~\cite{2016-Subramanian-Arxiv}. In the rest of this subsection, we obtain the PMF power spectrum for the slow-roll inflation. Appendix \ref{app:power-law} contains the results for power-law and de Sitter inflation.

To obtain the solution to Eq.~\eqref{eq:GFieldEOM} in slow-roll inflation, we introduce a new set of variables~\cite{2003-Martin.Schwarz-PRD,2004-Shankaranarayanan.Sriramkumar-PRD}:
\begin{equation}\label{new-var2}
x = \ln{\frac{aH}{k c_A}},~~
\mathcal{A}_k = e^{-\frac{x}{2}} (1-\epsilon_1)^{-\frac{1}{2}} u_k \, ,
\end{equation}
where $\epsilon_1$ is the first-order slow-roll parameter defined as:
\begin{equation}
\label{def:Slowroll}
    \epsilon_1 = -\frac{\dot{H}}{H^2} = 1 - \frac{\sH^{\p}}{\sH^2}
\end{equation}
Note that the above variables are well-defined up to the exit of inflation $(\epsilon_1 = 1)$. To solve the differential equation ~\eqref{eq:GFieldEOM}, we need to evaluate $Z$ in terms of the slow-roll parameters.  
Rewriting Eq.~\eqref{def:Slowroll}, we have~\cite{2001-Schwarz.etal-PLB},
\begin{equation}
\label{def:eta-epsilon}
\eta = - \frac{1}{(1 - \epsilon_1) \mathcal{H}} - \int \frac{2 \epsilon_1 \, 
(\epsilon_1 - \epsilon_2)}{(1 - \epsilon_1)^3} \, d\left(\frac{1}{\mathcal{H}}\right) \, ,
\end{equation}
where $\epsilon_2$ is the second slow-roll parameter \eqref{def:Slowrollpara}. Note that the second term in the above expression can be ignored when $\mathcal{H}$ is approximately constant and/or $\epsilon_1 \simeq \epsilon_2$. Under this condition, we have:
\begin{eqnarray} \label{slow-r}
aH \simeq -\frac{1}{\eta (1 - \epsilon_1)} ~~ \Longrightarrow~~x \simeq \ln{\frac{1}{c_A k \eta (\epsilon_1 - 1)}} \, .
\end{eqnarray}
We would like to note the following points: We do not assume $\epsilon_1 \ll 1$. $\eta$ is negative during inflation, hence $x$ is a well-defined. Using Eqs.~\eqref{new-var2} and \eqref{slow-r}, we have: 
\begin{eqnarray}
    \sH = e^x k c_A \, , &~~~~& 
    \sH^{\p} = e^{2x}k^2 (1-\epsilon_1) c_A^2 \, ,\\
    \sH^{\p\p} = 2 e^{3x}k^3 (1-\epsilon_1)^2 c_A^3\, , &~~~~& \sH^{\p\p\p} = 6 e^{4x}k^4(1-\epsilon_1)^3 c_A^4.
\end{eqnarray}

In the new variables, the adiabatic sound speed ($c_A$) and $Z''/Z$ become:
\begin{eqnarray}
\label{eq:cAslowroll}
c_A^2 &\simeq& 1 + (d_1-s_1) \frac{H}{\Lambda} + \left[s_1^2-s_2-s_1d_1+d_2+(e_1-b_1)
(1-\epsilon_1)\right] \left(\frac{H}{\Lambda}
\right)^2 \\
%%%
\frac{Z''}{Z} &\simeq&
 \frac{e^{2x} k^2}{4} \left\{2 s_1 \epsilon_1(\epsilon-1)
\frac{H}{\Lambda} \right. \nonumber \\
\label{eq:Zslowroll}
& + & \left.
\left[a_1^2 (2+2\epsilon_1-3 \epsilon_1^2) -a_2 (4-7\epsilon_1+\epsilon^2)-b_1 \epsilon_1
 (1-4\epsilon_1+3\epsilon_1^2)\right] \left(\frac{H}{\Lambda} \right)^2\right\} \, .
\end{eqnarray}
where we have truncated the series up to $(H/\Lambda)^2$ and we have not imposed any slow-roll approximation. Substituting the above expressions (\ref{new-var2}, \ref{eq:cAslowroll}, \ref{eq:Zslowroll}) in Eq. \eqref{eq:GFieldEOM} leads to: 
\begin{equation}
\label{eq:FinVecEq}
\frac{d^2 u_k}{dx^2} + \left[ q_1^2 e^{-2x} 
- q_2^2 \right] u_k =0 \, .
\end{equation}
where, 
{\small
\begin{eqnarray}
q_1 &=& (1-\epsilon_1)^{-1} \\
%%%%    
q_2 &=& \frac{1}{2}+ \left\{\frac{s_1
\epsilon_1 (1 - 2\epsilon_1)}{2}\frac{H}{\Lambda} - \left[\frac{s_1^2}{4}(2+2\epsilon_1-3 \epsilon_1^2) -s_2 (4-7\epsilon_1+\epsilon_1^2)-b_1 \epsilon_1(1-4\epsilon_1 - 3\epsilon_1^2)\right]\left(\frac{H}{\Lambda} \right)^2\right\} 
\nonumber \\
\label{q2-sr}
& & \times (1-\epsilon_1)^{-2} - \frac{s_1^2 \epsilon_1^2}{8} 
\frac{(1-2 \epsilon_1)^2}{(1-\epsilon_1)^{4}} 
\left(\frac{H}{\Lambda} \right)^2 
\end{eqnarray}
}
We want to note that we have not assumed $\epsilon_1 \ll 1$. Assuming $\epsilon_1$ is approximately constant, the solution to the above differential equation \eqref{eq:FinVecEq} is Hankel functions~\cite{abramowitz+stegun}
\begin{equation}
u_k(x) = \alpha \, H_{q_2}^{(1)}(e^{-x}q_1) 
+ \beta \, H_{q_2}^{(2)}(e^{-x}q_1) \, .
\end{equation}
Using the relation \eqref{new-var2}, the mode functions of the gauge field $\mathcal{A}_k$ are:
%
\iffalse
\begin{eqnarray}
    \mathcal{A} &=& (1-\epsilon_1)^{-\frac{1}{2}}
    \frac{(-q_1 k\eta)^{\frac{1}{2}}}{(1+\epsilon_1)^{\frac{1}{2}}}
    \left\{\alpha H_{q_2}^{(1)}(e^{-x}q_1) + \beta H_{q_2}^{(2)}(e^{-x}q_1)\right\}\\
  \mathcal{A}  &=& (-q_1k\eta)^{\frac{1}{2}}  
    \left\{\alpha H_{q_2}^{(1)}(e^{-x}q_1) + \beta H_{q_2}^{(2)}(e^{-x}q_1)\right\} ~~~
    \text{(first order in slow-roll)}.
\end{eqnarray}
\fi
%
\begin{equation}
\label{eq:EFT-Modefunction}
\mathcal{A}_k  = (-c_A k\eta)^{\frac{1}{2}}  
\left\{\alpha H_{q_2}^{(1)}(e^{-x}q_1) + \beta H_{q_2}^{(2)}(e^{-x}q_1)\right\}.
\end{equation}

In the sub-horizon limit ($c_A k |\eta| \gg 1$), we assume that the modes satisfy the Bunch-Davies vacuum, i. e., the mode approaches the Minkowski space behavior in the asymptotic past:
\begin{align}
\label{eq:BunchDaviesEFT}
\lim_{k\eta \to -\infty}  \mathcal{A}_k (\eta) =  \frac{1}{\sqrt{2c_A k}}e^{-ic_A k\eta}  
\, .
\end{align}
This leads to:
\begin{equation}
    \alpha = 0;~~~~ \beta = \sqrt{\frac{\pi}{4 c_A k}} \, .
\end{equation}
Substituting the mode function 
\eqref{eq:EFT-Modefunction}, 
with the Bunch-Davies vacuum initial condition,  in Eq.~\eqref{eq:pMS}, the 
power spectrum for the super-horizon modes ($c_A k |\eta| \ll 1$) is:
\begin{equation}
     \mathcal{P}_B=\frac{k^5}{a^4} \frac{c_A^2}{f_2}\frac{\eta}{8\pi^3}(\Gamma(q_2))^2
    \left(\frac{- c_A k\eta }{2}\right)^{-2 q_2}
\end{equation}
Substituting Eq. \eqref{slow-r}, 
we have:
\begin{equation}\label{PS-sr}
\mathcal{P}_B = \frac{H^4 (1-\epsilon_1)^4}{8 \pi^3} 
\frac{\Gamma^2(q_2)}{c_A^3 f_2}
(c_A k \eta)^{5-2 q_2} \, .
\end{equation}
By expanding $c_A$ and $f_2$ up to the second-order in $(\mathcal{H}/\Lambda)$, we have:
\begin{multline}\label{PS-srFin}
\mathcal{P}_B = H^4 (1-\epsilon_1)^4 
\Gamma^2(q_2) 
(c_A k \eta)^{5-2 q_2} \times \\
\left\{1+\frac{3 s_1-5d_1}{2}\frac{H}{\Lambda}+\frac{1}{2}[3 s_2-6s_1 d_1 -5d_2 +8d_1^2 +(3b_1-5e_1)(1-\epsilon)]\left(\frac{H}{\Lambda}\right)^2\right\} \, .
\end{multline}

This is the third key result of this work, regarding which we want to discuss the following points: 
First, the analysis is valid for all values of $\epsilon_1$, assuming that the contribution of the second term in Eq.~\eqref{def:eta-epsilon} can be ignored. While this is true during most of the inflation, the above expression may not be valid at the end of inflation.
Second, in Appendix \ref{app:power-law}, we have derived the power spectrum for de Sitter and Power-law inflation. While the power spectrum for the de Sitter inflation is exact, we have used the WKB approximation for the power-law case. Setting 
the $\epsilon_1 = 0$ in the above power spectrum \eqref{PS-sr} matches with the magnetic power spectrum \eqref{de-sitter-ps} during de Sitter. 
Third, from Eq.~(\ref{PS-srFin}), we see that the amplification in the magnetic power spectrum --- evaluated at the horizon crossing $(-k \eta c_A = 1$) ---  is possible whenever $3 s_1 - 5 d_1 > 0$.
Specifically, we see for sub-luminal ($d_1 < s_1$) or super-luminal ($s_1 < 0$) modes, the leading order correction term in the power-spectrum ${3 s_1-5d_1} > 0$.
In other words, we can have large amplifications even for super-luminal fluctuations. To avoid EFTs with superluminal fluctuations~\cite{2006-Adams.etal-JHEP}, we need another physical condition --- the modes should be sub-luminal.
Thus, our analysis clearly shows that magnetogenesis models must satisfy two necessary conditions --- conformal symmetry breaking and causal propagation. In the next section, we consider a specific model and show that this is the case.

\begin{figure}
\centering
%\subfigure[]{%
\label{b1-eps1}%
\includegraphics[height=2.85in]{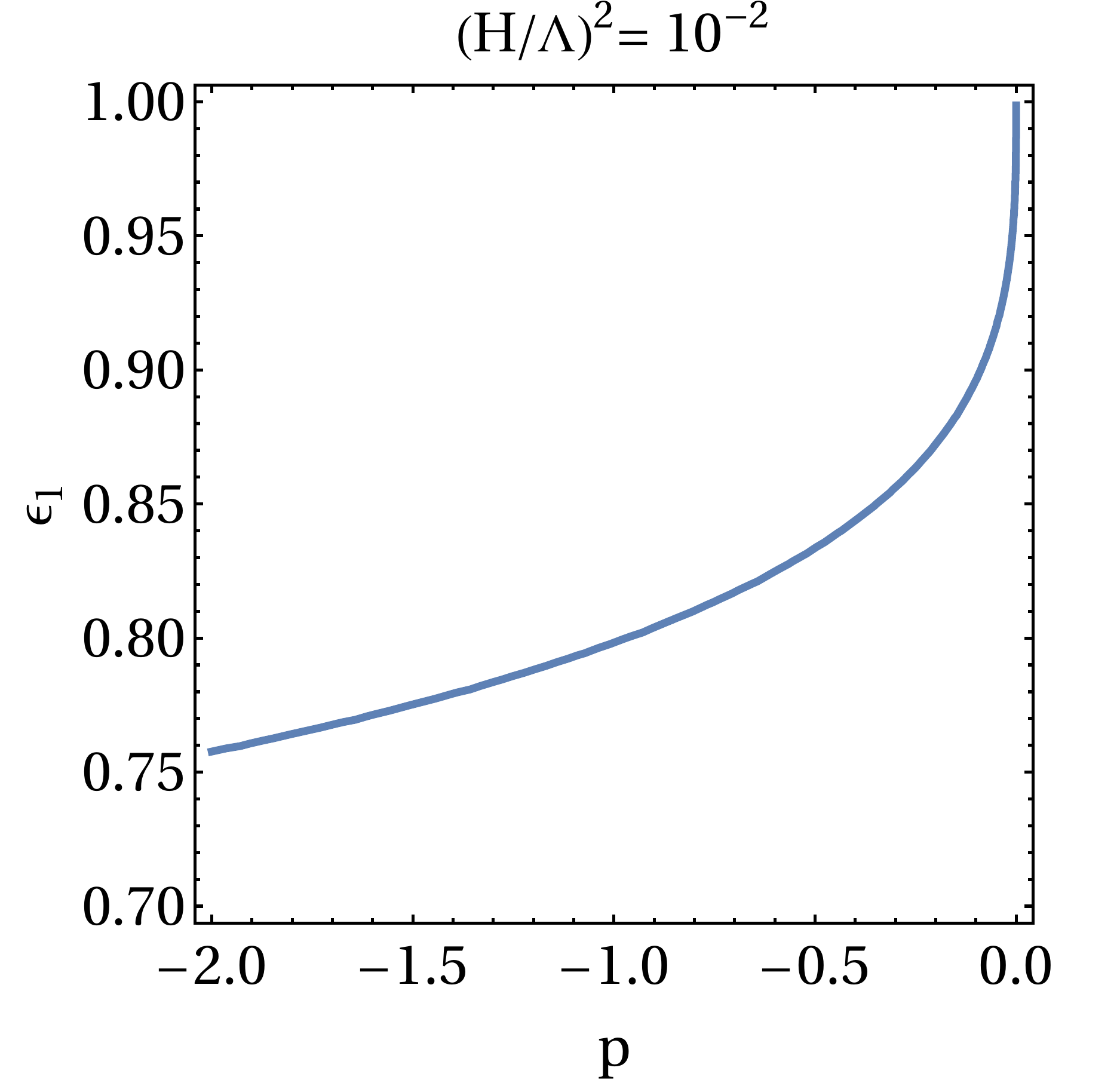}%}%
\quad
%\hspace{1.0cm}
%\subfigure[]{%
\label{b1-eps2}%
\includegraphics[height=2.85in]{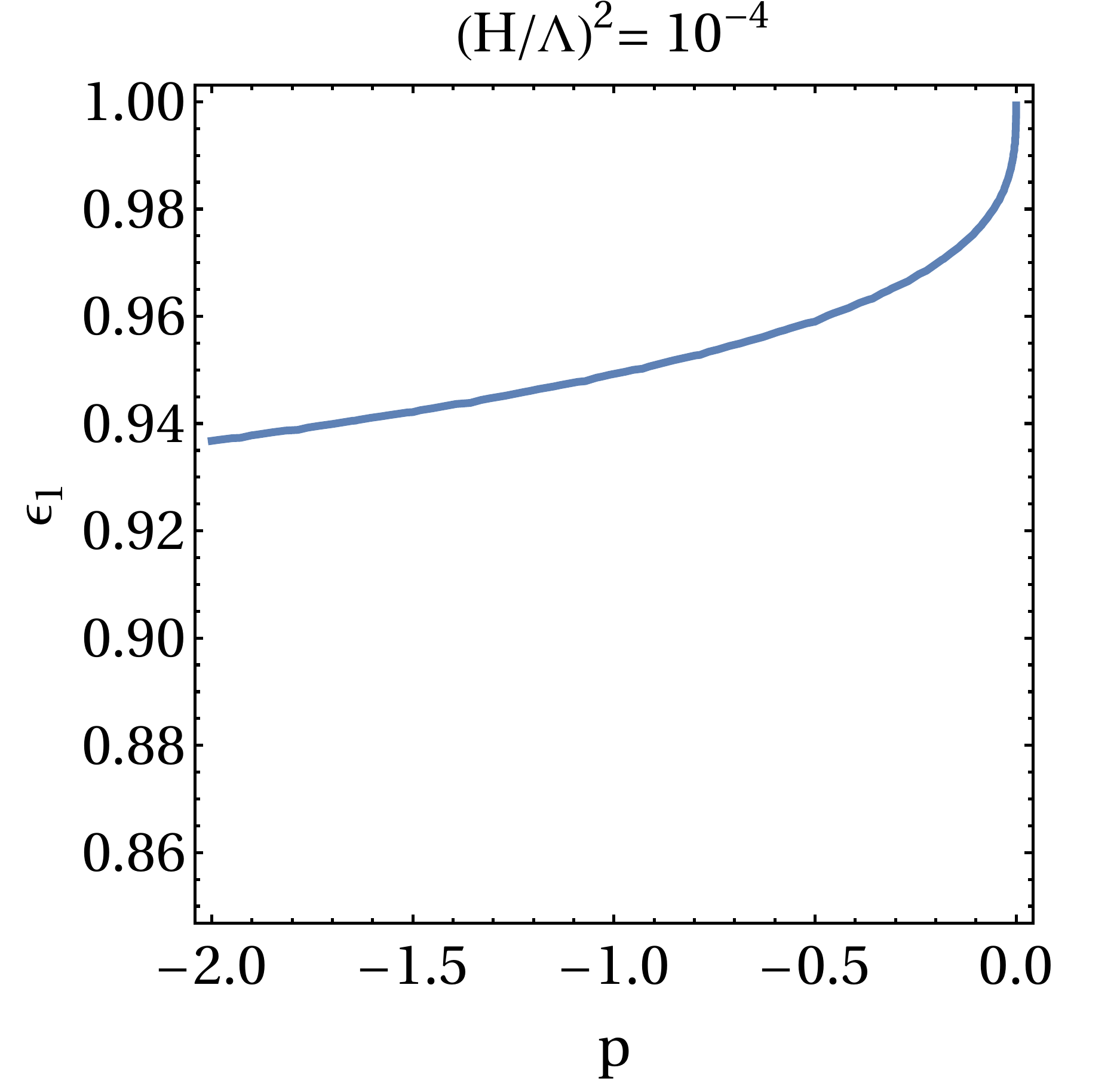}%}%
\caption{The allowed curves for two different values of $(H/\Lambda)^2$, for which we can get a scale invariant power spectrum for the magnetic field. The parameter $p$ is
$\mathcal{O}(1)$ and this curve is only valid at the end of inflation as
we have considered the value of the first slow roll parameter 
$\epsilon_1 \rightarrow 1$. In the left panel we have set $(H/\Lambda)^2 = 10^{-2}$ and in the right panel we have set
$(H/\Lambda)^2 = 10^{-4}$.}
\label{fig:b1-eps}
\end{figure}

Lastly, the above expression provides the following condition under which the magnetic field power spectrum \eqref{PS-sr} will be scale-invariant:
\begin{equation}\label{index-sr}
5-2 q_2 =0 \, .
\end{equation}
From Eq.~\eqref{q2-sr} we see that parameters $s_1$ and $s_2$ are associated with the $(H/\Lambda)$ expansion terms, while the parameter $b_1$ is associated with $\left(\sH^{\p}/a \Lambda\right)^2$.
While the terms containing the parameters $s_1$ and $s_2$ do not have time-dependent factors, the term containing $b_1$  has time-dependent factors. This can be seen by rewriting $\left(\sH^{\p}/a \Lambda\right)^2$ as $(1-\epsilon_1) \left(H/\Lambda \right)^2$. If we demand that the prefactors in the expansion are $\mathcal{O}(1)$, then this implies that $s_1, s_2$ 
and $((1-\epsilon_1) b_1)$ are all order unity during the entire inflationary epoch. In order for this to be satisfied, in general, $b_1$ can have a large value even with $\left(H/\Lambda\right)^2$ suppression. Hence, in order 
to get a scale-invariant power spectrum, the above condition reduces to:
\begin{equation}
\label{index-sr2}
4 - \frac{p \epsilon_1 (1-4 \epsilon_1 - 3 \epsilon_1^2)}{(1-\epsilon_1)^3} \left(\frac{H}{\Lambda}\right)^2 = 0 \,  
\quad {\rm where} \quad  p = b_1 (1 - \epsilon_1) \, .
\end{equation}
Note that the above redefinition ensures that $p$ is $\mathcal{O}(1)$ close to the end of inflation.  \ref{fig:b1-eps} contains the allowed values of $\epsilon_1$ for different values of $p$ for a fixed $({H}/{\Lambda})^2$. From the plot we see that for the scale invariant power spectrum in slow-roll scenario requires $b_1<0$.
 
 \section{Example for EFT Magnetogenesis: Vector Galileon model}
\label{sec:vector_galileon_example}
In the earlier section, using EFT, we showed that conformal invariance breaking is not a sufficient condition for generating the primordial magnetic field during inflation. In this section, we take a specific model and show that the generation of primordial magnetic fields require two necessary conditions --- conformal invariance breaking and causal propogation. 

In Ref.~\cite{2017-Debottam.Shankaranarayanan-JCAP}, two of the current authors have constructed a consistent vector Galileon model by demanding the following conditions in action: theory be described by vector potential $A_{\mu}$ and its derivatives, $U(1)$ gauge invariance should be preserved, and the equations of motion must be second order. The exciting feature of the model is that it only contains the derivative coupling --- higher derivative terms. Hence, it explicitly breaks the conformal invariance of the action, and therefore it may lead to the generation of magnetic fields in the early Universe.
Hence, this model has a clear advantage over other models, particularly the scalar field coupled models~\cite{1991-Ratra-Apj.Lett,2017-Sharma.etal-PRD}. Moreover, due to the absence of coupling between scalar field and electromagnetic field, the strong coupling problem is avoided naturally; hence, the model does not require fine-tuning.

In this section, we consider the total action, i.e., standard electromagnetic action with modification due to vector Galileon, and show that the conformal invariance breaking is the necessary condition to amplify the magnetic field generated during inflation, but it is not sufficient. More specifically, we show that the amplification of order ($\sim 10^7$) comes with a price that speed of sound $c_s>1$, without leading to the back-reaction problem. For $c_s<1$, the model generates tiny magnetic fields on large scales. Thus, despite having Lorentz-invariant Lagrangians, such theories admit super-luminal fluctuations can not be low-energy effective field theories of magnetogenesis.

\subsection{Generation of primordial magnetic fields}
\label{sec:PMF_generation}
This subsection will discuss the phenomenological consequences of the vector Galileon model in the early universe. The vector Galileon action is:
\begin{equation}
\label{def:VECaction}
 \mathcal{S}_{\rm{tot}} =  S_{\rm EM} + 
 \mathcal{S}_{\rm{VEC}} 
\end{equation}
where $S_{\rm EM}$ is the standard electromagnetic action given in Eq.~\eqref{eq:SEM} and $\mathcal{S}_{\rm{VEC}}$ is the vector Galileon part. In the flat 
FRW background (\ref{def:FRW}), with $N(\eta) = a(\eta)$, we have\footnote{ Note that in obtaining Eq.(\ref{eq:EMaction-FRW}) and Eq.(\ref{eq:TheModelFRW}), we have set $N(\eta) = a(\eta)$ in 
Eq.~(\ref{appeq:EMaction-FRW}) and 
Eq.~(\ref{appeq:TheModelFRW-simplify}), respectively} :
\begin{align}\label{eq:EMaction-FRW}
S_{EM} = \frac{1}{2} \int d^4x \, \left[  {A_i^{\prime}}^2 - (\partial_i A_j)^2  \right] \, ,
\end{align}

\begin{figure}
\centering
%\subfigure[]{%
\label{fig:csgt1}%
\includegraphics[height=2in]{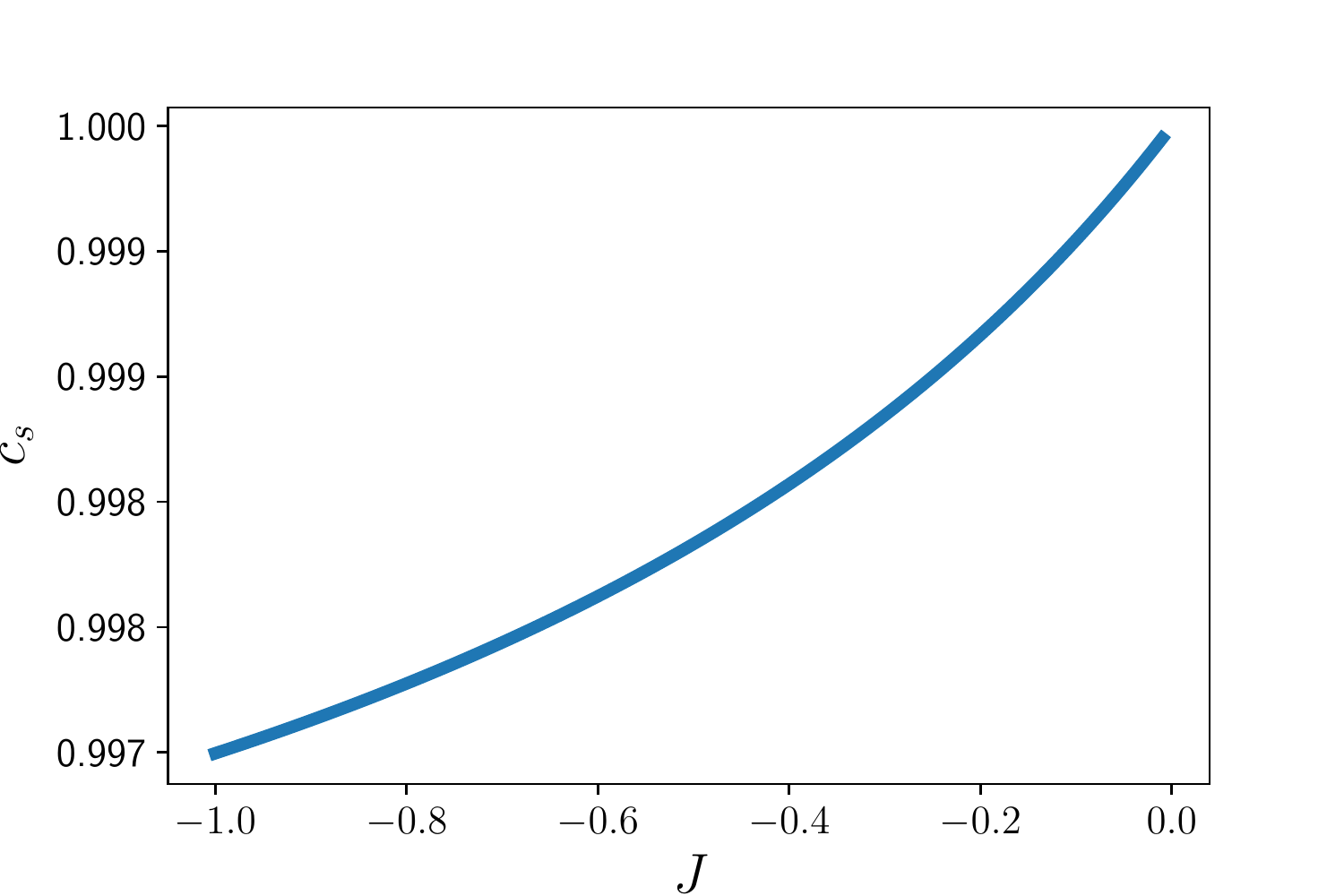}%}%
%\qquad
%\subfigure[]{%
\label{fig:cslt1}%
\includegraphics[height=2in]{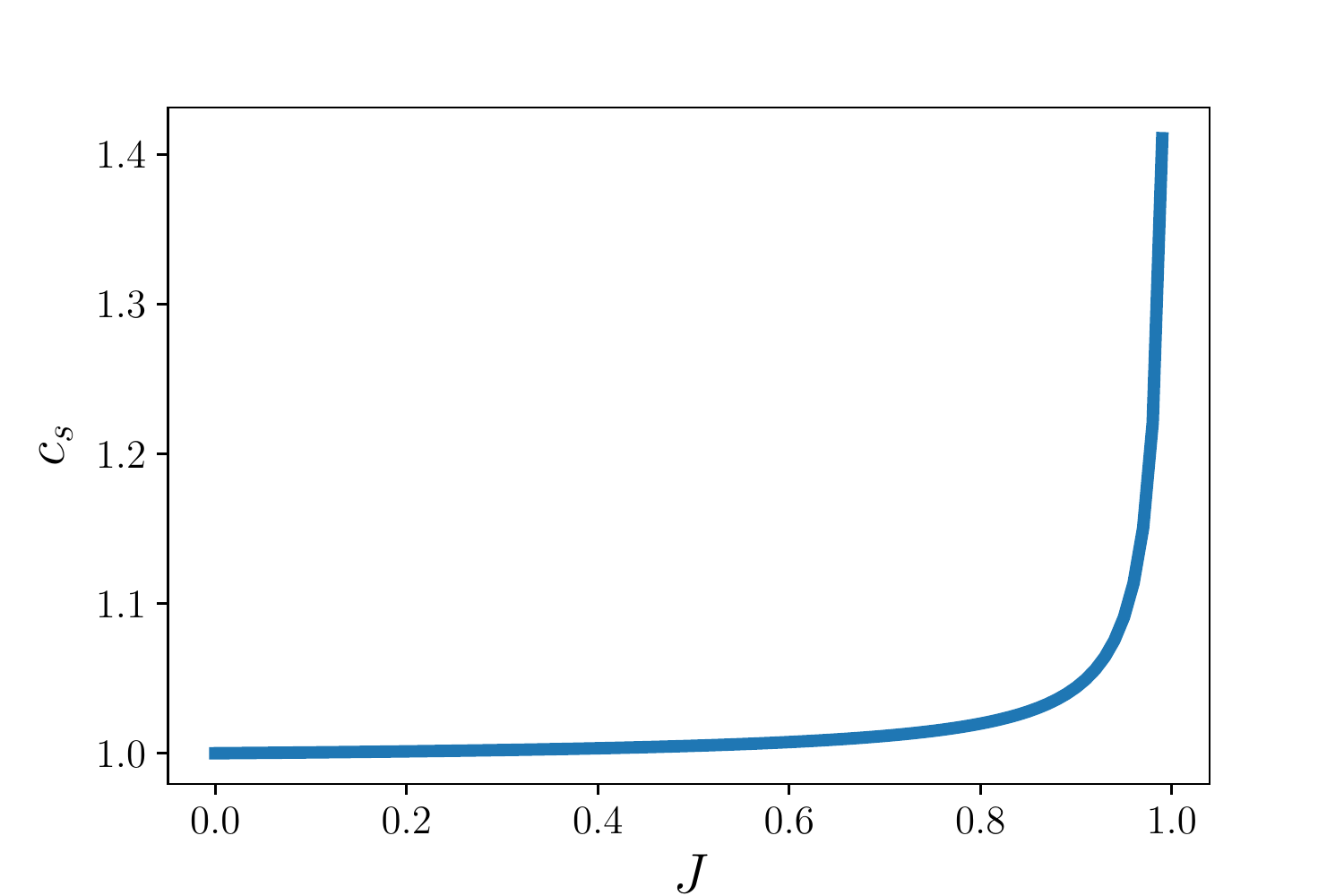}%}%
\caption{Plot showing the behaviour of $c_s$ with respect to parameter $J$.}
\label{fig:cs-plot}
\end{figure}

The vector Galileon action in FRW background is~\cite{2017-Debottam.Shankaranarayanan-JCAP}:
\begin{eqnarray}
\label{eq:TheModelFRW}
\mathcal{S}_{VEC} =  2D \, \int\,d^4x  \Big[- \,\frac{a^\prime{}^2}{a^4}\,A_i^\prime{}^2 +   \frac{a^{\prime\prime}}{a^3}\, \left(\partial_i A_j\right)^2  - \frac{{a^{\prime}}^2}{a^4}\,\left(\partial_i A_j\right)^2\Big] \, .
\end{eqnarray} 
where $A_{\mu}$ satisfies Coulomb gauge condition \eqref{def:CoulombG}. $D$ is the coupling constant with the dimensions of the square of the inverse of energy, i. e. $D \equiv 1/\Lambda^2$. We define the following dimensionless parameter $J$:
\begin{align}\label{eq:J-def}
J = 4 D H^2 = 4 \left(\frac{H}{\Lambda} \right)^2 \,  . 
\end{align}
Like the EFT action \eqref{EFT:L}, the 
above action (\ref{eq:TheModelFRW}) is not in canonical form; hence, we need to rewrite the action in canonical form. To do that, we define the canonical vector field $\mathbb{A}_i$ as:
\begin{align}\label{eq:def-canonicalA}
\mathbb{A}_i = \left( 1 - J \right)^{1/2} \, A_i \qquad \implies \qquad 
\left( 1 - J \right)^{1/2} \, A_i^{\prime} = \mathbb{A}_i^{\prime} + \frac{J ( {\mathcal{H}}^{\prime} - \mathcal{H}^2 )  }{ \mathcal{H} (1 - J ) } \mathbb{A}_i  \, .
\end{align}
Substituting Eqs. (\ref{eq:def-canonicalA}, \ref{eq:J-def}) in Eq.~\eqref{def:VECaction} and setting $N(\eta) = a(\eta)$ leads to: 
\begin{align}\label{eq:tot_action-canonical}
\mathcal{S}_{\rm{tot}} &=  \frac{1}{2} \, \int\,d^4x  \left[ \,\,\mathbb{A}_i^{\prime}{}^2  - \left[  \left(  \frac{\mathcal{H}^{\prime} - \mathcal{H}^2  }{\mathcal{H}}  \right)^2 \frac{J}{ (1 - J)^2 }   +   \frac{  J \left( \mathcal{H}^{\prime\prime} - 3 \mathcal{H}^{\prime} \mathcal{H} + \mathcal{H}^3 \right)  }{(1-J) \, \mathcal{H}}   \right]  \,\mathbb{A}_i^2 \right.
\nonumber \\
&{}{} \qquad  \left.  -  \left( 1 -  \frac{   J( \mathcal{H}^{\prime} - \mathcal{H}^2 ) }{(1-J) \mathcal{H}^2}  \right) \,   \left(\partial_i \mathbb{A}_j\right)^2 \right] \, ,
\end{align}
which is in the canonical form. The equation of motion for the canonical vector field can be obtained by varying the action (\ref{eq:tot_action-canonical}) with respect to $\mathcal{A}_i$:
{\small
\begin{align}\label{eq:eom-canonicalA}
\mathbb{A}_i^{\prime\prime} -  c_s^2 \, \nabla^2 \mathbb{A}_i  + \left[  \left(  \frac{\mathcal{H}^{\prime} - \mathcal{H}^2  }{\mathcal{H}}  \right)^2 \frac{J}{ (1 - J)^2 }   +   \frac{  J \left( \mathcal{H}^{\prime\prime} - 3 \mathcal{H}^{\prime} \mathcal{H} + \mathcal{H}^3 \right)  }{(1-J) \, \mathcal{H}}   \right] \,\mathbb{A}_i   = 0 \, 
\end{align}
}
where $c_s$ is the propagation speed of electromagnetic fluctuations and is given by:
\begin{align}\label{eq:def-cs}
c_s  = \sqrt{1 -  \frac{   J( \mathcal{H}^{\prime} - \mathcal{H}^2 ) }{(1-J) \mathcal{H}^2}} 
= \sqrt{\frac{1 - J (1 - \epsilon_1)}{ 1 - J} }  \, .
\end{align}
In obtaining the last expression, we have used the definition \eqref{def:Slowroll}. Substituting the Fourier decomposing of the canonical vector field
$\mathbb{A}_i$ (\ref{eq:fourierd}) in Eq.~\eqref{eq:eom-canonicalA}, we get:
{\small
\begin{align}\label{eq:eom-canonicalA-fourier}
\mathbb{A}_k^{\prime\prime} + \left[
c_s^2 \, k^2  + \left( \,\,  \left(  \frac{\mathcal{H}^{\prime} - \mathcal{H}^2  }{\mathcal{H}}  \right)^2 \frac{J}{ (1 - J)^2 }   +   \frac{  J \left( \mathcal{H}^{\prime\prime} - 3 \mathcal{H}^{\prime} \mathcal{H} + \mathcal{H}^3 \right)  }{(1-J) \, \mathcal{H}}   \right) \,\,  \right] \mathbb{A}_k = 0 \, .
\end{align}
}
In terms of the slow-roll parameters, the above expression reduces to: 
\begin{align}\label{eq:eom-SR-canonicalA}
\mathbb{A}_k^{\prime\prime} +  \left[ c_s^2 \, k^2 -    \frac{\mathcal{H}^2 J \epsilon_1}{(1 - J )^2} \, 
\left( [1 - J] (1  - 3\epsilon_1 + \epsilon_2)
- J\epsilon_1   \right) \right] \,\mathbb{A}_k = 0  \,  ,
\end{align}
where $\epsilon_1$ and $\epsilon_2$ are the slow-roll parameters defined in Eq.\eqref{def:Slowrollpara}. (See Appendix \ref{app:VG-slow-roll}.)
This is a key expression regarding which we would like to discuss the following: 
First, the above expression is exact and valid for any cosmological scenario. Although we have expressed the equation in terms of slow-roll parameters, we have not set $\epsilon_1 \ll 1$. Second, positive permittivity provides a condition on the value of $J$ to be $J < 1$. Hence, the transformation 
(\ref{eq:def-canonicalA}) is well-defined for all values of $J < 1$.
Third, in the limit of $J \to 0$, $c_s = 1$ and matches with standard electrodynamics. However, for any non-zero value of $J$, the sound speed is not unity; depending on the value of $J$, it can be greater or less than $1$. In the case of inflation, $\epsilon_1$ is positive, and hence, for the positive values of $J$, $c_s > 1$. However, in the case of super-inflation ($\dot{H}>0$)~\cite{2001-Gunzig.etal-PRD,2014-Basak-Shanki-JCAP}, $ \epsilon_1<0$ leading to the speed of sound less than 1~\cite{2014-Basak-Shanki-JCAP}. In the case of negative values of $J$, the sound speed is always less than unity. \ref{fig:cs-plot} contains the plot of $c_s$ as a function of $J$. It shows that $c_s > 1$ for positive $J$ during inflation.

Our next step is to solve Eq.~\eqref{eq:eom-SR-canonicalA}.
It is not possible to solve the above equation exactly for arbitrary $a(\eta)$. Hence, we consider two --- sub-horizon and super-horizon ---  limits of the above differential equation and match the mode functions at the sound horizon.
 
In the sub-horizon limit ($c_s  k | \eta| \gg 1$), we assume that the modes satisfy the Bunch-Davies vacuum, i. e., the mode approaches the Minkowski space behavior in the asymptotic past:
\begin{align}\label{eq:BunchDavies}
\lim_{k\eta \to -\infty}  \mathbb{A}_k (\eta) =  \frac{1}{\sqrt{2c_s k}}e^{-ic_s k\eta}  \,\,   .
\end{align}
For the physical gauge field ($A_{k}(\eta)$), using the relation (\ref{eq:def-canonicalA}), the above condition translates to: 
\begin{align}\label{eq:BunchDavis-ActualVec}
\lim_{k\eta \to -\infty} A_k (\eta) =  \frac{1}{\sqrt{2 k c_s \, (1 - J)}}  
%\, \left( \frac{1 - J ( 1 - \epsilon_{1})}{ 1 - J}  \right)^{-1/4}  \,
e^{-ik c_s \eta} 
%\, \sqrt{ \frac{1 - J + J \epsilon_{1}}{ 1 - J} } }
\end{align}
As we will see below, the above relation would be convenient to match the solution at the sound-horizon crossing to fix the arbitrary coefficients. In the super-horizon limit ($ c_s k | \eta | \ll 1$), we have 
\begin{align}\label{eq:superhor}
A_k (\eta) \simeq  C_1 +  C_2 \,\, \frac{ (1 - J)^{1 + \frac{2 J \epsilon_1 }{ 1 - J } } }{ 1 - J + 2 J \epsilon_1} \left( - k \eta \right)^{1 + \frac{2 J \epsilon_1 }{ 1 - J } }   \,\,  ,
\end{align}
where $C_1$ and $C_2$ are arbitrary coefficients. Since ${2 J \epsilon_1 }/(1 - J)$ is positive, and hence at the end of inflation, i.e., $\eta \rightarrow 0$, the second term in the above expression is negligible. In other words, the vector field freezes at the end of inflation and is approximately constant $C_1$.

As mentioned earlier, we demand that the sub-horizon modes \eqref{eq:BunchDavis-ActualVec} and super-horizon modes \eqref{eq:superhor} are continuous and differentiable at the sound-horizon. From Eq.(\ref{eq:eom-SR-canonicalA}), we obtain 
the horizon crossing time $\eta_*$ as
\begin{align}\label{eq:eta-cross}
\eta_*  = -  \frac{1}{k}\left[  \, \frac{ J_* \epsilon_1^* ( 1-J_* - \epsilon_1^*(3 - 2 J_*) + \epsilon_2^* (1-J_*) )   }{ (1-J_*) (1-J_* + J_* \epsilon_1^*) }   \, \right]^{1/2} \, ,
\end{align}
where $"*"$ refers to the quantities evaluated at the horizon crossing time $\eta_*$. Matching the mode functions at $\eta_*$ leads to: 
\begin{align}\label{eq:C1}
C_1 &= \frac{  ( 1 - J_* + 2 J_* \epsilon_1^* ) - i \sqrt{  J_* \epsilon_1^*\left( 1 - J_* - \epsilon_1^* (3 - 2 J_*) + \epsilon_2^*(1-J_*) \right)  } }{  \sqrt{2 c_s^* k (1 - J_*) }  \, (1 - J_* + 2 \epsilon_1^* J_*) }    e^{i \sqrt{ \frac{ J_* \epsilon_1^*\left( 1 - J_* - \epsilon_1^* (3 - 2 J_*) + \epsilon_2^*(1-J_*) \right) }{ (1-J_*)^2 } } }
   \\
\label{eq:C2}
C_2 &=   \frac{i \, {c_s^*}^{\frac{1}{2}+\frac{2J_* \epsilon_1^*}{1 - J_*}}}{\sqrt{2k(1-J_*)}} \left(  J_* \epsilon_1^*\left( 1 - J_* - \epsilon_1^* (3 - 2 J_*) + \epsilon_2^*(1-J_*) \right)  \right)^{\frac{ 2 J_*\epsilon_1^*}{J_* - 1}}  e^{i \sqrt{ \frac{ J_* \epsilon_1^*\left( 1 - J_* - \epsilon_1^* (3 - 2 J_*) + \epsilon_2^*(1-J_*) \right) }{ (1-J_*)^2 } } }
\end{align}

\subsection{Power spectrum and estimation of magnetic field strength}
\label{sec:power_spectrum}

To study the observable effects, we evaluate the energy
density of the electromagnetic fields~\cite{2013-Durrer.Neronov-Arxiv}. The $0-0$th component of the energy-momentum tensor $T_{\mu\nu}$ in the FRW background (\ref{def:FRW}) is
\begin{align}\label{eq:T00-def}
 T_{0 0} = - \frac{N^2}{a^3} \frac{\delta \mathcal{L}}{\delta N} \, .
\end{align}
Substituting the action \eqref{def:VECaction}
in the above expression and setting 
$N(\eta) = a(\eta)$, we get:
\begin{align}\label{eq:rho}
\rho 
= \frac{1}{2 a^4 } (1 - 3 J) \, \delta^{i j} A_i^\prime A_j^\prime 
+  \frac{1}{2 a^4 } ( 1 +  2 J ) \,  \delta^{i k} \delta^{j l}\, \partial_i A_j \,\partial_k A_l  + \frac{J}{H \,a^5}\, \delta^{i j}\, A_i{}^\prime\, \nabla^2 A_j \, .
\end{align}

The first term is the energy density of the Electric field $(\rho_E)$. 
The second and the third terms are the energy densities of the
magnetic field $(\rho_B)$ and $(\rho_{B.B^\prime})$, respectively.
Note that the first two terms decay as $a^{-4}$ while the last term decays as $a^{-5}$. Hence, one can ignore the last term in evaluating the energy density of electromagnetic fluctuations. As we will see below, imposing the condition that the energy density is always positive provides a condition on $J$.

Using the decomposition (\ref{eq:fourierd}), the electric and magnetic
part of the perturbation spectrum per logarithmic interval can be
written as:
\begin{eqnarray}\label{eq:SpectraB}
&& \mathcal{P}_{B}(k) \equiv  \frac{\mbox{d}}{\mbox{dln}k} \langle 0|\hat{\rho}_{B^2}|0 \rangle = \frac{ (1 + 2 J ) }{2 \pi^2} \frac{k^5}{a^4} \left|A_k\right|^2 \\
 \label{eq:SpectraE}
&&  \mathcal{P}_{E}(k) \equiv  \frac{\mbox{d}}{\mbox{dln}k} \langle0|\hat{\rho}_{E^2}|0\rangle =  \frac{ (1 - 3 J) }{ 2 \pi^2 } \frac{k^3}{a^4} \left|A_k^\prime\right|^2 \\
  \label{eq:SpectraBB}
&& \mathcal{P}_{_{B.B^\prime}}(k) \equiv  \frac{\mbox{d}}{\mbox{dln}k} \langle0|\hat{\rho}_{_{B.B^\prime}}|0\rangle = -\frac{J }{ 4 \pi^2 } \frac{k^5}{a^4 \mathcal{H}} ( A_k^\prime A_k^* + {A_k^\prime}^* A_k ) ~~~~.
 \end{eqnarray}
During most of the Universe's history,  the electrical conductivity of the Universe is high. Hence, the electric fields decay and do not contribute to the energy density. Hence, like in Sec. \eqref{sec:EFT-Spectrum}, we will concentrate on the magnetic field component\footnote{In some models, the electric power spectrum dominates over the magnetic part~\cite{2009-Demozzi.etal-JCAP}. In our case, the electric power spectrum does not contribute.}. 
Imposing the condition that energy density is non-negative implies that $J \geq -1/2$. Combining the earlier constraint, this implies $J$ in the range $[-0.5, 1)$ is well-defined.

Substituting Eq.~(\ref{eq:superhor}) in Eq.~(\ref{eq:SpectraB}) leads to:
\begin{eqnarray}\label{eq:PowerSpectraB}
&& \mathcal{P}_{B}(k) = \frac{H^4 (- k \, \eta_f)^4 }{4 \pi^2 } \frac{ (1+2J_f) ( 1 - J_* (2 - 5\epsilon_1^*) + J^2_* (1 - 5\epsilon_1^*)  )  }{c_A^* (1-J_*) (1-J_*+2J_*\epsilon_1^*)^2}
 \end{eqnarray}
where $\eta_f$ refers to the end of inflation and $J_f = 4 D H_f^2$.  

We now calculate the energy density of the generated electromagnetic fields by integrating the power spectrum over the Fourier modes $k_i < k < k_f$ where $k_i$ and $k_f$ refer to the initial and final modes leaving the horizon during inflation, i. e.:
\begin{align}\label{eq:rho_B}
    \rho_{B} = \int_{k_i}^{k_f} \mathcal{P}_{B}(k) \, \mbox{dln}k
    = \frac{H_f^4 (- k_f \, \eta_f)^4 }{16 \pi^2 } \frac{ (1+2J_f) ( 1- J_*(2-5\epsilon_1^*) + J^2_* (1 - 5\epsilon_1^*) )  }{c_A^* (1-J_*) (1-J_*+2J_*\epsilon_1^*)^2} \, .
\end{align}
We want to stress that the above expression is for a generic inflation model and depends on the value of $J_*$. 
In the rest of the section, we determine the range of $J_*$ for which the 
generated magnetic fields are sufficient for the observable large-scale magnetic fields such that the generated magnetic fields do not affect the background FRW metric. To avoid the back-reaction, the energy density of the generated fields must be less than the total background energy density during inflation. Assuming slow-roll inflation $\epsilon_1 \ll 1$, 
the back-reaction parameter ($\xi$) for the Galileon model is~\cite{2009-Demozzi.etal-JCAP,2016-Subramanian-Arxiv,2020-Talebian.etal-arXiv}:
\begin{align}\label{eq:back-reaction}
\xi \equiv \frac{\rho_{ \rm{B} } }{ \rho_{\rm{Inf}} } \approx \frac{  H_f^2  }{ 48 \pi^2 \, M_{ \rm{pl} }^2 } \, \frac{1}{(1 - \, J_* )}   <  1
\end{align}
where $\rho_{\rm{Inf}} = 3 H_f^2 M_{ \rm{pl} }^2$ is the background energy density during inflation and $\rho_B$ is given by Eq.(\ref{eq:rho_B}). As mentioned above, we assumed $\epsilon_1 \ll 1$ and set $-k_f \eta_f \sim 1$. From the above condition, we get the following constraint:
\begin{align}\label{eq:Jc_constraint}
\frac{1}{1 - J_*} \,   <    \, 2.25 \times 10^{13}
\end{align}
Combined with the earlier constraint of $J$, we see that $J_*$ close to $1$ can lead to large amplification. Let us now consider two cases:
\begin{enumerate}
\item[${\bf J_* > 0}$:] When $J_*$ is very close to 1, we can have large amplification in the magnetic field strength.  
Let us now estimate the magnetic field strength corresponding to $(1 - J_*)^{-1}  \approx  2.25 \times 10^{13}$ at the comoving wavenumber $k = 1~{\rm Mpc}^{-1}$.  The magnetic field for a mode with wave vector $k$ at the end of inflation is given by:
\begin{align}\label{eq:Bf-exp}
B_f \approx \sqrt{ \frac{ H_f^4 k^4 \eta_f^4 }{4 \pi^2 c_s^* (1-J_*) } }  \approx 1.58 \times 10^{13}  \, \rm{G} \, ,
\end{align}
where we have set $| \eta_f | = 10^{-20} \rm{Mpc}$~\cite{2014-Martin.etal-PhyDarkUniv} and since $J_* \to 1$ we have approximated $c_s^* \simeq 1$. 

In the radiation-dominated (or matter-dominated) epoch, the magnetic field decays adiabatically as $B \propto a^{-2}$. We then have the following relation between the magnetic field strength at present and the end of inflation:
\begin{align}\label{eq:B0-Bf-rel}
B_0 = B_f  \left( \frac{a_f}{a_0}  \right)^2 .
\end{align}
Assuming the instantaneous reheating and the entropy conservation, i.e., $g T^3 a^3 = \rm{constant}$ during its evolution, where $T$ is the temperature of the relativistic fluid, and $g$ is the number of effective relativistic degrees of freedom~\cite{2016-Subramanian-Arxiv} gives
\begin{align}\label{eq:a0/af-rel}
\frac{a_0}{a_f} =  \left( \frac{90}{8\pi^3}  \right)^{1/4}  \frac{g_f^{1/12} }{g_0^{1/3} } \frac{\sqrt{ H_f \, M_{ \rm{pl} } } }{T_0} 
\approx 0.9 \times 10^{29} \, \left(  \frac{H_f}{10^{-5}  \, M_{ \rm{pl} }  }  \right)^{1/2}
\end{align}
where we have taken $g_f \approx 100$ and $g_0 = 2.64$. Substituting Eq.~(\ref{eq:a0/af-rel}) in Eq.~(\ref{eq:B0-Bf-rel}),the present day magnetic field strength at $\rm{Mpc}^{-1}$ scale is:
\begin{align}\label{eq:B0estimate}
B_0 = 1.5 \times 10^{-45} \, \rm{G}
\end{align}
\item[${\bf J_* < 0}$:] In this case, $1/(1 - J_*)  \approx 1$. The magnetic field for a mode with wave vector $k$ at the end of inflation is given by:
\begin{align}\label{eq:Bf-exp2}
B_f \approx \sqrt{ \frac{ H_f^4 k^4 \eta_f^4 }{4 \pi^2 c_s^* } }  \approx 1.58 \times 10^{7}  \, \rm{G} \, ,
\end{align}
where, here again, we have set $| \eta_f | = 10^{-20} \rm{Mpc}$~\cite{2014-Martin.etal-PhyDarkUniv} and $c_s^* \simeq 1$. Following the same analysis, the present day magnetic field strength at $\rm{Mpc}^{-1}$ scale is:
\begin{align}\label{eq:B0estimate2}
B_0 = 1.5 \times 10^{-51} \, \rm{G}
\end{align}
\end{enumerate}
In estimating the magnetic field strength, we 
have used the standard values of inflation, however, if inflation occurs over the energy scales such that $10^{-10} \lesssim H_f / M_{ \rm{pl} } \lesssim 10^{-5} $~\cite{2021-Tripathy.etal-arXiv}, then the adiabatic expansion rate will be much less, i. e.,
\begin{align}
10^{26} \lesssim \frac{a_0}{a_f} \lesssim 10^{29}~~.
\end{align}
In that case, we gain six orders of magnitude in the magnetic field strength. 

Our analysis shows that appreciable magnetic field strength can be achieved when $J_* \to 1$. However, as can be see in \eqref{fig:cs-plot}, $J_* > 0$ leads to super-luminal electromagnetic perturbations. Let us now compare the above result with the EFT analysis. Following the discussion in Appendix \eqref{app:VGModel}, we 
find that $s_2$ and $e_1$ are the non-zero parameters that describe vector Galileon model. Substituting these in Eq.~\eqref{eq:cAslowroll}, we have:
\begin{equation}
  c_A = 1 - \frac{1}{2}\left[s_2 - e_1 (1-\epsilon_1)\right] \left(\frac{H}{\Lambda}\right)^2.
\end{equation}
Using the relation \eqref{eq:VG-EFTComp}, we have:
\begin{equation}
c_A = 1 + \frac{J \epsilon_1}{8} \,  .
\end{equation}
The EFT analysis shows that $J < 0$ avoids super-luminal propagation. Hence, we need another physical condition --- the modes should be sub-luminal. This conclusion is identical to the one we obtained from the EFT analysis. Thus, our analysis shows that conformal invariance breaking is necessary but not sufficient condition to generate sufficient magnetic fields in the early Universe.

Another interesting thing to note is that by setting the above parameters in the EFT power spectrum \eqref{PS-sr}, we can see that the vector Galilean model will never lead to a scale-invariant power spectrum and is consistent with the analysis in this section. 

\section{Conclusions and Discussions}
\label{sec:discussion}

The origin of primordial magnetic fields is still unresolved and requires physics beyond the standard models of cosmology and particle physics. Although inflation provides a causal mechanism for the generation of primordial density perturbations, it can not generate the appreciable primordial magnetic field in the early Universe. It has been argued that conformal invariance breaking is a sufficient condition to generate primordial magnetic fields during inflation. In this work, using EFT based on expansion about the Hubble parameter $(H)$ and its derivatives, we show that the generation of primordial magnetic fields requires two
necessary conditions --- conformal invariance breaking and causal propagation. We have also shown that a broad class of magnetogenesis models can be reproduced from the EFT that is a sum of series in $H/\Lambda$, time-derivatives of $H$ --- ($H'/\Lambda^2$), ($H''/\Lambda^3$), $\cdots$ --- and their products, for instance, $H H'/\Lambda^3$, $H H''/\Lambda^4$, $\cdots$.

Like the EFT of inflation, the EFT
of magnetogenesis requires the inclusion of fluctuations in the matter and metric degrees
of freedom. As shown in Appendix A, the gauge field is the relevant gauge-invariant variable for the EFT. Demanding that the EFT of magnetogenesis breaks conformal invariance, however, satisfies local Lorenz invariance and gauge invariance, we obtained the general second-order EFT action. This action depends on two expansion scalar functions --- $f_1(H, a, \Lambda)$ and $f_2(H, a, \Lambda)$. We then showed that the EFT action \eqref{EFT:L} could reproduce all the known magnetogenesis models. 
To make the computation of the power-spectrum tractable and to highlight the importance of speed of perturbations, we truncate the series \eqref{eft:A} up to second order. However, the truncation of the series to compute the power spectrum has no bearing on the EFT expansion \eqref{eft:A}.

By truncating the expansion scalar functions to $\Lambda^{-2}$, we derived the EOM of the gauge field and obtained the power spectrum in the slow-roll inflation scenario. From Eq.~(\ref{PS-srFin}), we see that the amplification in the magnetic power spectrum --- evaluated at the horizon crossing $(-k \eta c_A = 1$) ---  is possible whenever $3 s_1 - 5 d_1 > 0$. Specifically, we see for sub-luminal ($d_1 < s_1$) or super-luminal ($s_1 < 0$) modes, the leading order correction term in the power-spectrum ${3 s_1-5d_1} > 0$.
In other words, we can have large amplifications even for super-luminal fluctuations. To avoid EFTs with superluminal fluctuations~\cite{2006-Adams.etal-JHEP}, we need another physical condition --- the modes should be sub-luminal.

We then considered a specific model of inflationary magnetogensis where the vector Galileon breaks the conformal invariance. We extensively studied the magnetic field generation during inflation, considering the total action. Due to the absence of coupling between scalar field and electromagnetic field, the model does not lead to a strong coupling problem~\cite{2017-Debottam.Shankaranarayanan-JCAP}; hence, the model does not require fine-tuning, which is an interesting feature of our model.
Furthermore, the evolution of vector modes is frozen on the super-horizon scale. We showed that the model predicts the present-day magnetic field of strength $B_0 \sim 10^{-45} {\rm G}$ on the cosmological (Mpc) scale which is several orders higher compared to the standard electromagnetic action, and there is no back-reaction problem. For low-scale inflationary models i.e., $10^3 - 10^4 \, {\rm GeV}$, the model generates $B_0 \sim 10^{-33} \rm{G}$. However, this comes with the price --- super-luminal electromagnetic perturbations. In other words, the model generates large magnetic fields and does not violate any other conditions; it leads to super-luminal perturbations. 

Modifications of general relativity provide an alternative explanation for cosmological inflation~\cite{2022-Shankaranarayanan-Joseph-GRG}. 
Modified gravity theories have extra degrees of freedom that might have interesting physical consequences in the early Universe. For instance, Stelle gravity~\cite{1978-Stelle-GRG} contains massive tensor modes, and these modes carry more energy than the scalar modes in $f(R)$ gravity models~\cite{Chowdhury:2022ktf}. While the current analysis can be extended to $f(R)$ gravity models, it is not straightforward to extend to Stelle gravity. Investigating the EFT of inflation and magnetogenesis in these modified gravity models is interesting. Also, it will be interesting to extend the EFT of magnetogenesis to bouncing models. We hope to address these in the future. The current analysis can be extended to helical fields which is currently under investigation.

%===================== A P P E N D I X ===================
%
\appendix

\section{Decoupling of vector perturbations and gauge fields}
\label{sec:Decoupling}

As mentioned in the Introduction, we aim to construct an EFT of magnetogenesis. Like any other effective theory, the EFT of magnetogenesis includes two components: Symmetries and degrees of freedom~\cite{2007-Burgess-ARNPS,2020-Penco-Arxiv}. Thus, like any perturbation theory in gravity, the EFT of magnetogenesis requires the inclusion of fluctuations in the matter and metric degrees of freedom. As mentioned above, in the EFT of inflation, one makes a specific gauge choice in which the inflaton fluctuations vanish~\cite{2007-Cheung.etal-JHEP}. To consistently investigate magnetogenesis, we need to include the vector perturbations in the metric~\cite{2004-Battefeld.Brandenberger-PRD,2004-Lewis-PRD,2014-Basak-Shanki-JCAP}. In this appendix, we show that the vector perturbations do not have temporal evolution; hence, only the gauge field is the relevant gauge-invariant variable for the EFT.

We consider vector perturbations about a spatially flat $(3+1)-$dimensional FRW line element:
\begin{equation}
ds^2 = \overline{g}_{\mu\nu} dx^{\mu} dx^{\nu} = 
N^2(t) dt^2 - a^2(t) \delta_{ij} dx^i dx^j = 
N^2(\eta) \, d \eta^2 - a(\eta)^2\delta_{ij} dx^i dx^j \, .
\label{def:FRW}
\end{equation}
The only degree of freedom in the metric is the usual expansion parameter $a(t)$, a function only of time by virtue of the homogeneity of the three-space. The lapse function $N(t)$ simply represents the time reparametrization invariance.

The conformal time ($\eta$), 
is related to the cosmic time by the relation $t =\int d\eta a(\eta)$, $a(\eta)$ is the scale factor and $N(\eta) = a(\eta)$. $H$ is the
Hubble parameter given by $H \equiv \dot{a}/{a}$ while $\sH \equiv a'/a$ is related to the Hubble parameter by the relation $\sH = H\,a$.

At the linear level, for a single scalar field inflation with gauge-field $(A_{\mu})$, the metric perturbations ($\delta g_{\mu\nu}$) can be categorized into three distinct types — scalar, vector and tensor perturbations~\cite{1984-Kodama.Sasaki-PTPS,1992-Mukhanov.etal-Phy.Rep.}. As mentioned above, we will consider the gauge field perturbations for magnetogenesis. Hence, the perturbed FRW line-element corresponding to vector perturbations can be written as~\cite{1984-Kodama.Sasaki-PTPS,1992-Mukhanov.etal-Phy.Rep.,2004-Battefeld.Brandenberger-PRD},
\begin{equation}
\label{eq:VectPert}
ds^2 
= a^2(\eta)\left[ d\eta^2 + 2 S_i d \eta dx^i - \left(\delta_{ij}+ 2 F_{(i,j)} \right) \, dx^i dx^j \right] \, ,
\end{equation}
where $S_i$ and $F_i$ are pure vector contributions and are divergenceless~\cite{1984-Kodama.Sasaki-PTPS,1992-Mukhanov.etal-Phy.Rep.}:
\begin{equation}
    \partial_i S^i = 0; ~~
    \partial_i F^i = 0;~~F_{(i,j)}=\frac{F_{i,j}+F_{j,i}}{2} \, .
\end{equation}
The above-perturbed metric has four unknown vector-mode functions. Since there are only two physically relevant vector modes, we must make certain choices on $S_i$ and $F_i$. We choose~\cite{2004-Battefeld.Brandenberger-PRD}:
\begin{equation}
    F_i = 0.
\end{equation}

To keep calculations transparent, let us consider the following action~\cite{2021-Giovannini-PLB}: 
\begin{equation}\label{test-L}
\mathcal{S}= S_{\rm EH} + S_{\rm Inflaton} + S_{\rm EM} + S_{\rm EM-NM}
\end{equation}
where,
\begin{eqnarray}
S_{\rm EH} &=& - \int d^4x \sqrt{-g} \frac{M_{\rm Pl}^2}{2} R \, , \nonumber \\
%%%%
S_{\rm Inflaton} &=& \int d^4x \sqrt{-g} \left[ \frac{1}{2}\partial_{\mu} \phi \partial ^{\mu} \phi - V(\phi)  \right], \nonumber \\
\label{def:EMaction}
S_{\rm EM} &=& - \frac{1}{4} 
\int d^4x \sqrt{-g} \,  
F_{\mu \nu} F^{\mu \nu} 
\, , \\
%%%
\label{def:EM-Nonminimal}
S_{\rm EM-NM} &=& \frac{1}{M^2}\int d^4x \sqrt{-g}\left[\Lambda_1(\varphi)R F_{\mu \nu} F^{\mu \nu} +
\Lambda_2(\varphi) R^{\nu}_{\mu}F_{\nu \alpha} F^{\mu \alpha}+ \Lambda_3(\varphi)R_{\mu\nu\alpha\beta}F^{\mu\nu}F^{\alpha\beta}+\right.  \\ 
& & \left. 
+ \Lambda_4(\varphi)C_{\mu\nu\alpha\beta}F^{\mu\nu}F^{\alpha\beta}+ 
 \Lambda_5(\varphi)
\Box \varphi F_{\mu \nu} F^{\mu \nu} +
\Lambda_6(\varphi)\partial_{\mu}\varphi
\partial^{\nu}\varphi F^{\mu \alpha}F_{\nu \alpha}
+ \Lambda_7(\varphi)
\nabla_{\mu}\nabla^{\nu}\varphi F_{\nu \alpha} F^{\mu \alpha}\right] \nonumber
\end{eqnarray}

$\phi$ is the canonical scalar (inflaton) field, $V(\phi)$ is the self-interacting potential, $\varphi = \phi/M$ with $M$ being some characteristic scale,
$F_{\mu \nu} \equiv \partial_{\mu} A_{\nu} - \partial_{\nu} A_{\mu}$ is the electromagnetic tensor. Eq.~\eqref{def:EMaction} is the standard gauge-field action while action \eqref{def:EM-Nonminimal} corresponds to all possible non-minimal coupling terms of the gauge field with gravity and the inflaton~\cite{2021-Giovannini-PLB}. Owing to the fact that the background space-time is FRW, the natural choice for the \emph{classical} gauge field in the background space-time is $\bar{A}_{\mu} = 0$. Splitting the gauge-field in the background and perturbations \eqref{eq:VectPert}, we have:
\begin{equation}
    A_{\mu} = \bar{A}_{\mu} + \delta A_{\mu} = \delta A_{\mu}.
\end{equation}
Like scalar perturbations, gauge field perturbations are quantum fluctuations. 
In terms of the Stewart-Walker lemma, the quantities which vanish in the FRW universe are gauge invariant
perturbations~\cite{1974-Stewart-Walker-PRSLA}. 
In this case, it is easy to check that $\delta A_{\mu}$ is gauge invariant. As mentioned above, there are only two physically relevant vector modes on the metric side. This implies that the gauge field should have only two physically relevant modes. One of the simplest and most well-known choice is the Coulomb gauge, i. e., 
\begin{equation}
\delta A_0 = 0, \partial_i \delta A^{i}=0 \, .
\label{def:CoulombG}
\end{equation}

With the above choices for the metric and gauge field, expanding action \eqref{test-L} to second order in $S_i$ and $A_{\mu}$, we have:
\begin{equation}\label{vec-L}
    \mathcal{S}^{(2)} = \frac{M_{\rm Pl}^2}{4}
    \int d^4x a^2  \left(\partial_j S^i\right)^2 +
    \frac{1}{2} \int d^4x \, \left[g_1(a,H,M){\delta A_i^{\prime}}^2 - \partial_j (g_2(a,H,M)\delta A_i)^2  \right] \, ,
\end{equation}

where $g_1(a,H,t,M)$ and $g_2(a,H,t,M)$ are functions of scale factor 
$a$, Hubble parameter $H$. In the case of minimal coupling, $g_1, g_2$ become unity.
This is a crucial result, and the EFT we develop in this work depends on this, regarding which we would like to discuss the following points: First, 
the vector perturbation $S_i$ does not have any temporal evolution, while the gauge fields have temporal evolution even if one considers non-minimal couplings of the  gauge field with gravity and inflaton field as in
\eqref{def:EM-Nonminimal}. Since $S_i$ does not evolve in time, it does not affect the dynamics of the gauge field. Thus, $S_i$ can be considered an irrelevant degree of freedom. Second, 
the above action does not contain any coupling between $\delta A_i$ and $S_i$. Hence, the vector perturbation $S_i$ does not affect the dynamical evolution of the gauge field $\delta A_{i}$. This is even more interesting considering that the analysis is also valid for non-minimal coupling. Even though initially, the energy density of the gauge field may be negligible compared to the inflaton energy, due to non-minimal coupling, the energy density of the gauge field will evolve non-trivially during
the inflationary phase. 
Consequently, non-minimal coupling terms may cause isocurvature density perturbations~\cite{2005-Shankaranarayanan.Lubo-PRD}, and 
the gauge field can not be treated as a spectator field. However, from Eq. \eqref{vec-L}, we see that the vector modes do not evolve in time, and the only relevant degree of freedom is the gauge field.  See Ref.~\cite{Kubota:2022pit} for a similar study in the context of inflationary paradigm with a spectator field with non-minimal coupling. Third, the above analysis can be extended to the case where the 3-space is not flat. In that case, also $S_i$ does not evolve in time. Lastly, we can extend the computation for $f(R)$ theories as in second order in
perturbation $S_i$ will not have any time evolution for this kind of theory.
This conclusion can also be drawn for 
Horava-Lifshitz gravity \cite{2010-Gong.etal-PRD}. 
%Fifth, the above analysis can be extended to the bouncing scenario~\cite{2020-Nandi-PLB}, implying that the effective degrees of freedom for the vector perturbations is the gauge field $A_i$.

Lastly, constructing EFT magnetogenesis requires consistently counting the number of degrees of freedom to write down the EFT Lagrangian. Hence, we need to find any possible interaction between the vector modes of perturbation $S_i$ and $\delta A_i$. From \eqref{test-L}, one can see that the lowest order
interaction between $S_i$ and $\delta A_i$ would come from the standard electromagnetic term
$F_{\mu\nu}F^{\mu\nu}$ and is:
\begin{equation}
    F_{\mu \nu}F^{\mu \nu} = 2 S^i g^{jk} \, \partial_0 \delta A_j \, 
    \left(\partial_i \delta A_k - \partial_k \delta A_j\right) \, .
\end{equation}
As can be seen, this interaction term is third-order. Similarly, the non-minimal terms in Eq. \eqref{def:EM-Nonminimal} can also produce higher order interaction between gauge field
and vector modes.
While these are important for higher-order computations where one also has to include scalar and tensor perturbations~\cite{2010-Christopherson.etal-PRD}; for the current work, these terms are irrelevant. Hence, the interaction between $S_i$ and $\delta A_i$ can be ignored for the EFT of magnetogenesis containing quadratic terms, and the only relevant degrees of freedom will be the gauge field $A_{\mu}$.

\section{Correspondence between EFT parameters and magnetogenesis models}\label{app:coup-details}

In this appendix, we provide detailed calculations that show a one-to-one mapping between the magnetogenesis model and EFT parameters defined in Eq.~\eqref{eft:A}. Specifically, we consider 
Ratra model~\cite{1991-Ratra-Apj.Lett}, gravitational coupling of the electromagnetic field~\cite{1988-Turner.Widrow-PRD,2020-Kushwaha.Shankaranarayanan-PRD}, higher-order gravitational coupling~\cite{2022-Bertolami.etal-arXiv} and Galileon vector fields~\cite{2017-Debottam.Shankaranarayanan-JCAP}.

\subsection{Class of Ratra Model with integer exponent}

The Lagrangian corresponding to the conformal invariance breaking term in the Ratra model is~\cite{1991-Ratra-Apj.Lett}:
\begin{equation}
\mathcal{S}_{\rm Ratra} = \int d^4x \sqrt{-g} \, f^2(\phi)F_{\mu\nu}F^{\mu\nu} 
\end{equation}
where $f(\phi)$ is an arbitrary function of the inflaton field. 
While there have been many simple choices, here, we consider one choice~\cite{2012-Barnaby.Namba.Peloso-PRD}:
\begin{equation}
\label{def:fphi}
f(\phi) = f_0 \exp{\left(-\frac{C \phi^2}{\Lambda^2}\right)},
\end{equation}
where $f_0, C$ are positive constants. Note that in the limit of $\Lambda \to \infty$, $f(\phi) \to f_0$ and is consistent with the series expansion~\eqref{eft:A}.

For the inflationary potential~\cite{Chervon:2017kgn}
\begin{equation}
V(\phi) = -\frac{8}{3} M^2_{\rm Pl} \lambda_2 \phi^2 + \lambda_2 \phi^4 \, ,
\end{equation}
the scalar field $\phi$ evolves as:
\begin{equation}
    \phi^2 = \sqrt{3} M_{\rm Pl} H
\end{equation}
Substituting the above expression in Eq. \eqref{def:fphi}, we have:
\begin{equation}
    f(\phi) = f_0 \exp{\left(-\sqrt{3} \tilde{C} \frac{H}{\Lambda}\right)} \qquad \tilde{C} = C \frac{M_{\rm Pl}}{\Lambda}
\end{equation}
Expanding the above form of $f(\phi)$ implies that the series will only consist of integer powers of $H$ and can be easily mapped to the EFT action \eqref{EFT:L}.
%Thus, the Ratra model for the above form of $f(\phi)$ correspond to the terms 
%$\left(\frac{\sH}{\Lambda}\right)^{C \sqrt{\frac{p}{2}}}$ term  in the EFT action \eqref{EFT:L}. Similarly, for other forms of $f(\phi)$ one can obtain a one-to-one mapping with the EFT parameters. 
Recently, in Ref.~\cite{2022-Durrer.etal-arXiv}, authors have considered magnetogenesis in the Higgs-Starobinsky model.  This leads to a generalized form of $f(\phi)$ in the Einstein frame.  By repeating the 
same procedure, we can obtain a one-to-one mapping with the EFT parameters $s_n, d_n$.

%\subsubsection{Slow-roll Inflation}
%\textcolor{blue}{
%For slow roll inflation the inflaton field value can be estimated as,
%\begin{equation}
%    V(\phi) \sim \frac{1}{8 \pi G} (3 H^2 + \dot{H})
%\end{equation}
%So for potentials of the type of $\lambda_n \phi^n$ (where $\lambda_n $ is some
%coupling constant)we can have,%
%\begin{equation}
%    \phi \sim \frac{1}{8 \pi G \lambda_n} (3 H^2 + \dot{H})^{\frac{1}{n}}
%\end{equation}
%Now if we consider a conformal symmetry breaking coupling function for magnetogenesis %as,
%\begin{equation}
%    f(\phi) \sim (\phi / \phi_0)^m, 
%\end{equation}
%then we can see that for a choice of $(m/n) \rightarrow$ integer, the model can be 
%mapped to our EFT expansion. 
%Other choices of potential can also be mapped to our EFT expansion for appropriate 
%choices of parameters.}

\subsection{Gravitational Coupling}

The Lagrangian corresponding to the non-minimal coupling of the electromagnetic field tensor with curvature terms are~\cite{1988-Turner.Widrow-PRD}:
\begin{equation}\label{gr-coup}
\mathcal{S}_{\rm NC} =\frac{1}{4}\int d^4 x \left(- \Gamma_1 R F_{\mu\nu}F^{\mu\nu}
    - \Gamma_2 R_{\mu\nu}g_{\alpha \beta}F^{\mu\alpha}F^{\nu\beta}
    - \Gamma_3 R_{\mu\nu\alpha\beta} F^{\mu\nu}F^{\alpha\beta}\right)
\end{equation}
where $\Gamma_1, \Gamma_2$ and $\Gamma_3$ are coupling constants. For the FRW background \eqref{def:FRW}, the Ricci scalar ($R$) is: 
\begin{equation}
\label{FRW:Ricciscalar}
    R = -6 \left( \frac{\sH^2+\sH^{\p}}{a^2}\right).
\end{equation}
Substituting in the term in Eq. \eqref{gr-coup}, we can see that the conformal invariance breaking terms correspond to 
$s_2 =b_1 = d_2 =  e_1 = -3$ in Eq.~\eqref{eft:A}. Let us now focus on the 
second term ($R_{\mu\nu}g_{\alpha \beta}F^{\mu\alpha}F^{\nu\beta}$). For the FRW background, we have: 
\begin{equation}
   R_{\mu\nu}g_{\alpha \beta}F^{\mu\alpha}F^{\nu\beta}=
   -\frac{1}{a^2}\left(\frac{\sH^2}{2}+\sH^{\p} \right)
   (A_i^{\p})^2+ 
   \frac{1}{a^2} \left(\frac{\sH^{\p}}{2}+\sH^2\right)
    (\partial_i A_j)^2
\end{equation}
Comparing the above expression with Eq.~\eqref{eft:A}, we see that the conformal invariance breaking terms correspond to $s_2=-\frac{1}{2}, b_1 =-1$ 
and $d_2 =-1, e_1=-\frac{1}{2}$. Finally, expanding $R_{\mu\nu\alpha\beta} F^{\mu\nu}F^{\alpha\beta}$ term for the FRW background, we have: 
\begin{equation}
    R_{\mu\nu\alpha\beta} F^{\mu\nu}F^{\alpha\beta} = 
    -\frac{\sH^{\p}}{a^2}(A_i^{\p})^2 + \frac{\sH^2}{a^2} (\partial_i A_j)^2.
\end{equation}
Here again, choosing $b_1=d_1=-1$ in \eqref{eft:A} corresponds to the Riemann coupling terms in the above action 
\eqref{gr-coup}.

\subsection{Higher Order Gravitational Coupling}

The Lagrangian corresponding to the non-minimal coupling of the electromagnetic field tensor with higher-order Ricci scalar is~\cite{2022-Bertolami.etal-arXiv}:
\begin{equation}
\mathcal{S}_{\rm Higher} 
=\int d^4x \sqrt{-g} \, R^3 F_{\mu\nu}F^{\mu\nu}
\end{equation}
Substituting the Ricci scalar \eqref{FRW:Ricciscalar} in the above expression, it can be shown that the above 
action has a one-to-one correspondence with EFT action \eqref{eft:A}, for the following choice of parameters: 
\[
s_6 = -6, b_3 = -6, g_{41} = -18, g_{22} = 
-18, d_6 = -6, e_3 = -6, h_{41} = -18, h_{22} = -18 \, .
\]

\subsection{Vector Galileon Model}
\label{app:VGModel}

Let us now consider the conformal invariance breaking term in the vector Galileon model~\cite{2017-Debottam.Shankaranarayanan-JCAP}:
 \begin{equation}
\mathcal{S}_{\rm VG} =  2 \, D 
\int d^4 x \left[-\frac{a'^2}{N^3 a}(A_i')^2 + \left(\frac{a''}{N a^2} - \frac{a' N'}{N^2 a^2}\right)(\partial_i A_j)^2 \right] 
\label{Galileon}
\end{equation}
where for the FRW metric \eqref{def:FRW} the lapse function $N = a(\eta)$. Comparing 
\eqref{Galileon} and \eqref{eft:A}, 
we see that all parameters except $s_2$  and $e_1$ vanish in \eqref{eft:A}. We thus get: 
\begin{equation}
\label{eq:VG-EFTComp}
s_2 = e_1, ~{\rm and}~D = - \frac{s_2}{\Lambda^2} \, .
\end{equation}
%========  Energy Density for EFT ================

%
\section{EFT action with different coupling functions for the components of the gauge field}\label{general-propagation}
In general, we can have different coefficients in front of components of the gauge field $(A_1, A_2, A_3)$ in the EFT action~(\ref{EFT:L}). However, as shown in this appendix, this will lead to different propagation speeds for these three components. To see this, we consider the following action:
\begin{equation}\label{general-coeff}
\mathcal{S} = \int d^4 x \left\lbrace\left( f_{1} \, A_1^{\prime 2} + f_{1,2} \, A_2^{\prime 2} + f_{1,3} \, A_3^{\prime 2}\right) - \left( f_{2} \, (\partial_j A_1)^2 + f_{2,2} \,(\partial_j A_2)^2 + f_{2,3} \, (\partial_j A_3)^2\right)\right\rbrace.
\end{equation}
First, we can introduce the following field redefinitions of the gauge field components $(A_2, A_3)$:
\begin{equation}\label{redef}
\tilde{A}_2^{\prime} = \left(\frac{f_{1,2}}{f_1}\right)^{1/2} A_2^{\prime} ; \quad 
\tilde{A}_3^{\prime} = \left(\frac{f_{1,3}}{f_1}\right)^{1/2} A_3^{\prime}  \, .
\end{equation}
Thus, \eqref{general-coeff} becomes:  
\begin{equation}\label{general-coeff2}
\mathcal{S} = \int d^4 x \left\lbrace f_{1} \left( A_1^{\prime 2} +  \tilde{A}_2^{\prime 2} + \tilde{A}_3^{\prime 2}\right) - \left( f_{2} \, (\partial_j A_1)^2 + f_{2,2} \,(\partial_j A_2)^2 + f_{2,3} \, (\partial_j A_3)^2\right)\right\rbrace.
\end{equation}
Here we can notice that the first three terms have the same coupling functions as $f_1$, but the action is written in terms of $A_1$, $A_2$, $A_3$ and $\Tilde{A}_2$ and $\Tilde{A}_3$. To write down the action in terms of $A_1$, $\Tilde{A}_2$ and $\Tilde{A}_3$ we next focus on the single component $A_2$ in \eqref{general-coeff2} which can be written in terms of $\Tilde{A}_2$ as,
\begin{equation}
    A_2 = \left(\frac{f_{1}}{f_{1,2}}\right)^{1/2} \tilde{A}_2 - \int d\eta \, \frac{d}{d \eta}\left(\frac{f_{1}}{f_{1,2}}\right)^{1/2} \tilde{A}_2(\eta)
\end{equation}
then $f_{2,2} \,(\partial_j A_2)^2$ term in \eqref{general-coeff2} becomes:
\begin{align}
    f_{2,2} \,(\partial_j A_2)^2 &= f_{2,2} \, \left(\frac{f_{1}}{f_{1,2}}\right) (\partial_j \tilde{A}_2)^2 + f_{2,2}\int d\eta \, d\eta^{\p} \, \frac{d}{d \eta}\left(\frac{f_{1}}{f_{1,2}}\right)^{1/2} \partial_j \tilde{A}_2(\eta) \frac{d}{d \eta^{\p}}\left(\frac{f_{1}}{f_{1,2}}\right)^{1/2} \partial_j \tilde{A}_2(\eta^{\p}) \nonumber \\ 
    &{} - 2 f_{2,2} \, \left(\frac{f_{1}}{f_{1,2}}\right)^{1/2} \partial_j \tilde{A}_2 \int d\eta \, \frac{d}{d \eta}\left(\frac{f_{1}}{f_{1,2}}\right)^{1/2} \partial_j \tilde{A}_2(\eta)\\
    &= F_{2,2} (\partial_j \tilde{A}_2)^2
\end{align}
where, $F_{2,2}$ is a function of $f_1$, $f_{1,2}$, $f_{2,2}$ and $\eta$. A similar exercise can be done for the $A_3$ component to express it in terms of $\Tilde{A}_3$, and we can introduce a new coupling function as $F_{2,3}$. Hence, after the field redefinition \eqref{general-coeff2} can be written as,
\begin{equation}\label{general-coeff3}
\mathcal{S} = \int d^4 x \left\lbrace f_{1} \left( A_1^{\prime 2} +  A_2^{\prime 2} + A_3^{\prime 2}\right) - \left( f_{2} \, (\partial_j A_1)^2 + F_{2,2} \,(\partial_j A_2)^2 + F_{2,3} \, (\partial_j A_3)^2\right)\right\rbrace.
\end{equation}
Here we have omitted the "tilde" from the fields. We can also notice that even if the temporal derivatives of the field components have the same coupling function, the spatial derivative part of the field components have different coupling functions. Hence, the propagation speeds of the three components are:
\begin{equation}
c_{A_1} = \left(\frac{f_{2}}{f_{1}}\right)^{1/2} ; \quad     \tilde{c}_{A_2} = \frac{F_{2,2}}{f_1} ; \quad
     \tilde{c}_{A_3} = \frac{F_{2,3}}{f_1}
\end{equation}
Hence, the propagation speed of the different components is different. The analysis in this work can be extended for this case.
%
%\begin{equation}
%    c_{A_1} = \left(\frac{f_{2}}{f_{1}}\right)^{1/2} ; \quad
 %   c_{A_2} = \left(\frac{f_{2,2}}{f_{1,2}}\right)^{1/2} ; \quad
 %   c_{A_3} = \left(\frac{f_{2,3}}{f_{1,3}}\right)^{1/2}
%\end{equation}

\section{Energy density for EFT action in terms of Lapse function}
\label{app:rhoEFT}
Vector perturbations can affect the background metric and matter configuration in which the perturbations propagate. This has been extensively studied for density perturbations during inflation and is also operational for other perturbations like vector perturbations. In order to quantify the effect of these EFT perturbations on the FRW background, we need to obtain the energy density ($\rho_{\rm EFT}$) corresponding to the EFT action \eqref{EFT:L}. 
%========

 To obtain the expression for the energy density, it is convenient to use the ADM formalism which simplifies the calculations significantly. The metric in ADM formalism is given by~\cite{1992-Mukhanov.etal-Phy.Rep.}:
 \begin{align}\label{appeq:ADM}
     ds^2 = N^2 d\eta^2 - \gamma_{ij} (N^i d\eta + dx^i) (N^j d\eta + dx^j)
 \end{align}
 where $N$ is lapse function, $N^i$ is shift vector and $\gamma_{ij}$ is the metric on constant $\eta-$hypersurface. Note that with $N^i =0$ and $\gamma_{ij} = a^2(\eta) \delta_{ij}$, we obtain flat FRW metric (\ref{def:FRW}). 
In the flat 
FRW background (\ref{def:FRW}), for arbitrary $N(\eta)$, the standard electromagnetic action becomes
\begin{align}\label{appeq:EMaction-FRW}
 S_{EM} = \frac{1}{2} \int d^4x \, \left[ \frac{a}{ N} {A_i^{\prime}}^2 - \frac{N}{a}(\partial_i A_j)^2  \right] \, ,
\end{align}
and the vector Galileon action (ignoring the total derivative term) is ~\cite{2017-Debottam.Shankaranarayanan-JCAP}:
%
%
\begin{comment}
\begin{eqnarray}
\label{appeq:TheModelFRW}
 \mathcal{S}_{VEC} =  2D \, \int\,d^4x  \Big[- \,\frac{a^\prime{}^2}{N^3\, a}\,A_i^\prime{}^2 +   \frac{a^{\prime\prime}}{N\,a^2}\, \left(\partial_i A_j\right)^2  - \frac{a^{\prime}\,N^\prime}{N^2\,a^2}\,\left(\partial_i A_j\right)^2\Big] \, .
\end{eqnarray} 
%
%$0-0$th component  of the energy-momentum tensor ($T_{\mu\nu}$) corresponding to the EFT action \eqref{EFT:L} in the FRW background \eqref{def:FRW} is
%
%\begin{align}\label{eq:energy_density-EFT}
%T_{00} = - \frac{N^2}{a^3} \frac{\delta \mathcal{S}_{\rm EFT} }{\delta N}
%\end{align}
%
Using the following relation:
%
\begin{align}
    \frac{\partial}{\partial\eta} \left[ \frac{a^{\prime}}{N a^2} \left(\partial_i A_j\right)^2 \right] = \left( \frac{a^{\prime\prime}}{N\,a^2}  - \frac{a^{\prime}\,N^\prime}{N^2\,a^2}\,\right)\left(\partial_i A_j\right)^2 - \frac{2 {a^{\prime}}^2}{N\,a^3}\, \left(\partial_i A_j\right)^2  + \frac{2 a^{\prime}}{N\,a^2}\,\partial_i A_j \, \partial_i A_j^{\prime}
\end{align}
%
in Eq.(\ref{appeq:TheModelFRW}) and ignoring the total derivative term (i.e., RHS of Eq.(\ref{appeq:term_simplify})), we obtain:
\end{comment}
%
\begin{eqnarray}
\label{appeq:TheModelFRW-simplify}
 \mathcal{S}_{VEC} =  2D \, \int\,d^4x  \Big[- \,\frac{a^\prime{}^2}{N^3\, a}\,A_i^\prime{}^2 +   \frac{2 {a^{\prime}}^2}{N\,a^3}\, \left(\partial_i A_j\right)^2  - \frac{2 a^{\prime}}{N\,a^2}\,\partial_i A_j \, \partial_i A_j^{\prime}\Big] \, .
\end{eqnarray} 
The lapse function does not have a kinetic term, and the variation of the total (gravity and matter) action w.r.t $N$ will lead to a constraint. Rewriting the total action 
in the form 
\begin{equation}
S_{\rm Grav-Mat} = S_{\rm EH} + S_{\rm tot} = \int d^3x a^3 \int dt \left[ p_a \dot{a} + p_{A_i} \dot{A}_i - N \, H_{A_i} \right] \, ,
\end{equation}
where $p_a$ is the canonical momentum corresponding to $a$, $p_{A_i}$ is the canonical momentum corresponding to $A_i$ and $H_{A_i}$ is the Hamiltonian corresponding to the EM field, including the Vector Galileon term.

\begin{comment}
Now with these settings, we can calculate the $0-0$th component  of the energy-momentum tensor ($T_{\mu\nu}$) as :
%
\begin{align}\label{eq:energy_density-EFT-1}
T_{00} = - \frac{N^2}{a^3} \frac{\delta \mathcal{S} }{\delta N},
\end{align}
%
and the energy density corresponding to the action is given by:
%
\begin{align}\label{appeq:rho-def-1}
\rho = T^0_0 = g^{00} T_{00} = - \frac{1}{a^3} \frac{\delta \mathcal{S} }{\delta N} \, .
    \end{align}
%
\end{comment}

We can now extend the above analysis for the EFT action \eqref{EFT:L}. In other words,  the EFT action can be written as:
\begin{equation}
\mathcal{S}_{\rm EFT} = \int d^3x a^3 \int dt \left[  p_{A_i} \dot{A}_i - N \, H_{\rm EFT} \right] \, ,
\end{equation}
where $H_{\rm EFT} = \int d^3x a^3 \rho_{\rm EFT}$ is the Hamiltonian 
corresponding to the EM field with EFT expansion. The corresponding energy density is:
\begin{align}\label{appeq:rho-def}
\rho_{\rm EFT} = T^0_0 = g^{00} T_{00} = - \frac{1}{a^3} \frac{\delta \mathcal{S}_{\rm EFT} }{\delta N} \, .
\end{align}
Note that in order to use the above method,
it is convenient to rewrite the expansion parameter in terms of the lapse function in the FRW line element (\ref{def:FRW}).
However, when we wrote the expansion parameters in Eq.~(\ref{eft:A}) we have set $N(\eta) = a(\eta)$. Hence, we first need to reinstate the lapse function $N(\eta)$ in Eq.~(\ref{eft:A}). This can be done in two steps: 
\begin{enumerate}
   \item[Step 1:] We first write the EFT parameters $f_i(\mathcal{H}, a, \eta)$, defined in Eq.~(\ref{eft:A}) (for $i = 1,2$), in terms of the cosmic time:  
   \begin{align}
    f_i(\mathcal{H}, a, \eta)  
    \xrightarrow{a\, d\eta \, \rightarrow \, dt}
    f_i(H, a, t)   
    \end{align}
\item[Step 2:] Rewrite the EFT parameters in cosmic time $f_i(H,a,t)$ in terms of conformal time and arbitrary lapse function $N(\eta)$:  
    \begin{align}
    f_i(H, a, t)   
    \xrightarrow{dt \, \rightarrow \, N d\eta}
    f_i(\mathcal{H}, a, N, \eta) 
    \end{align}
\end{enumerate}    
We need to be careful to obtain $f_i(\mathcal{H}, a, N, \eta)$,  as the zeroth order terms (i.e., $s_0, d_0$) in Eq.~(\ref{eft:A}) are independent of $N(\eta)$. 
%a(\eta)$ in the setting $N(\eta) = a(\eta)$(which is used in Eq.(\ref{eft:A}) ). 
We fix these zeroth-order terms in the EFT expansion parameters with respect to the standard electromagnetic action for the metric (\ref{def:FRW}): 
    \begin{align}
        S_{\rm EM} = -\frac{1}{4} \int d^4x \sqrt{-g} F_{\mu\nu} F^{\mu\nu} = \int d^4x \left[ \frac{a(\eta)}{2N(\eta)} (A_i^{\prime})^2 - \frac{N(\eta)}{2a(\eta)} (\partial_i A_j)^2 \right]
    \end{align}
Thus, by fixing the zeroth-order EFT parameters with $S_{\rm EM}$ as $a(\eta) s_0/2N(\eta)$ and $N(\eta) d_0/2a(\eta)$, we will obtain the expansion parameters $f_1$ and $f_2$. Note that in the setting $N(\eta) = a(\eta)$ we get the zeroth order terms as $s_0, d_0$ (cf. Eq.(\ref{eft:truncated})), which are time-independent and can be set to unity. 

Following these two steps, the expansion parameters ($f_1, f_2$), in terms of lapse function, take the following forms:
\begin{equation}
\label{appeq:EP-lapse}
\begin{split}
 &  f_1(\mathcal{H}, a, N, \eta) = 
    \frac{a}{N} s_0 + \sum_{n=1}^{\infty} \frac{s_n}{N^n}  \left(\frac{\sH}{\Lambda}\right)^n
    + \sum_{n=1}^{\infty} \frac{b_n}{a} \frac{1}{N^n}
    \left(\frac{1}{\Lambda^2} \frac{d}{d\eta} \left( \frac{a}{N} \mathcal{H} \right) \right)^n+...
\\
& f_2(\mathcal{H}, a, N, \eta) = 
    \frac{N}{a} d_0 + \sum_{n=1}^{\infty} \frac{d_n}{N^n}  \left(\frac{\sH}{\Lambda}\right)^n
    + \sum_{n=1}^{\infty} \frac{e_n}{a} \frac{1}{N^n}
    \left(\frac{1}{\Lambda^2} \frac{d}{d\eta} \left( \frac{a}{N} \mathcal{H} \right) \right)^n+...
\end{split}
\end{equation}
As mentioned above, we have separated the zeroth-order terms in the expansion parameters and are fixed with respect to the standard electrodynamics action. 
Setting $N(\eta) = a(\eta)$ in the above expression matches with Eq.(\ref{eft:A}). 
Using the expansion parameters (\ref{appeq:EP-lapse}) in terms of lapse function and varying the action (\ref{eft:A}) with respect to the lapse function $N \rightarrow N+\delta N$ gives:
\begin{align}\label{appeq:deltaS1}
    \delta \mathcal{S}_{\rm EFT} &= - \int d^4x \, \left[ \left( \frac{a}{N^2} s_0 + \sum \frac{n}{N^{n+1}} s_n \left( \frac{\mathcal{H}}{\Lambda} \right)^n + \sum \frac{n}{a N^{n+1} \Lambda^{2n}} b_n \left( \frac{d}{d\eta}\left( \frac{a \mathcal{H}}{N} \right) \right)^n\right) \delta N \, (A_i^{\prime})^2 
    \right. \nonumber\\
    &{} \left. 
    - (A_i^{\prime})^2 \sum \frac{n}{a N^n \Lambda^{2n}} b_n \left( \frac{d}{d\eta}\left( \frac{a \mathcal{H}}{N} \right) \right)^{n-1} \frac{d}{d\eta}\left( \frac{a \mathcal{H}}{N^2} \delta N \right) + .....
    \right. \nonumber\\
    &{} \left.
    + \left( \frac{1}{a} d_0 - \sum \frac{n}{N^{n+1}} d_n \left( \frac{\mathcal{H}}{\Lambda} \right)^n - \sum \frac{n}{a N^{n+1} \Lambda^{2n}} e_n \left( \frac{d}{d\eta}\left( \frac{a \mathcal{H}}{N} \right) \right)^n\right) \delta N \, (\partial_i A_j)^2  \right. \nonumber\\
    &{} \left. 
    + (\partial_i A_j)^2 \sum \frac{n}{a N^n \Lambda^{2n}} e_n \left( \frac{d}{d\eta}\left( \frac{a \mathcal{H}}{N} \right) \right)^{n-1} \frac{d}{d\eta}\left( \frac{a \mathcal{H}}{N^2} \delta N \right) + .....
    \right].
\end{align}
Doing integration by parts in the second and fourth line in the equation (\ref{appeq:deltaS1})
and ignoring the total derivative terms gives:
\begin{align}\label{appeq:deltaS}
    \delta \mathcal{S}_{\rm EFT} &= - \int d^4x \, \left[ \left( \frac{a}{N^2} s_0 + \sum \frac{n}{N^{n+1}} s_n \left( \frac{\mathcal{H}}{\Lambda} \right)^n + \sum \frac{n}{a N^{n+1} \Lambda^{2n}} b_n \left( \frac{d}{d\eta}\left( \frac{a \mathcal{H}}{N} \right) \right)^n\right) \, (A_i^{\prime})^2 
    \right. \nonumber\\
    &{} \left. 
    +  \frac{a \mathcal{H}}{N^2} \frac{d}{d\eta} \left( (A_i^{\prime})^2 \sum \frac{n}{a N^n \Lambda^{2n}} b_n \left( \frac{d}{d\eta}\left( \frac{a \mathcal{H}}{N} \right) \right)^{n-1}  \right) + .....
    \right. \nonumber\\
    &{} \left.
    + \left( \frac{1}{a} d_0 - \sum \frac{n}{N^{n+1}} d_n \left( \frac{\mathcal{H}}{\Lambda} \right)^n - \sum \frac{n}{a N^{n+1} \Lambda^{2n}} e_n \left( \frac{d}{d\eta}\left( \frac{a \mathcal{H}}{N} \right) \right)^n\right) \, (\partial_i A_j)^2  \right. \nonumber\\
    &{} \left. 
    - \frac{a \mathcal{H}}{N^2} \frac{d}{d\eta} \left( (\partial_i A_j)^2 \sum \frac{n}{a N^n \Lambda^{2n}} e_n \left( \frac{d}{d\eta}\left( \frac{a \mathcal{H}}{N} \right) \right)^{n-1}  \right) + .....
    \right]\delta N .
\end{align}
Now using Eq.~(\ref{appeq:rho-def}) and Eq.~(\ref{appeq:deltaS}), the energy density corresponding to the EFT action is:
\begin{align}\label{appeq:rhoEFT}
    \rho &= \frac{1}{a^4}\left[ \left( \frac{a^2}{N^2} s_0 + \sum \frac{n a}{N^{n+1}} s_n \left( \frac{\mathcal{H}}{\Lambda} \right)^n + \sum \frac{n}{ N^{n+1} \Lambda^{2n}} b_n \left( \frac{d}{d\eta}\left( \frac{a \mathcal{H}}{N} \right) \right)^n\right) \, (A_i^{\prime})^2 
    \right. \nonumber\\
    &{} \left. 
    +  \frac{a^2 \mathcal{H}}{N^2} \frac{d}{d\eta} \left( (A_i^{\prime})^2 \sum \frac{n}{a N^n \Lambda^{2n}} b_n \left( \frac{d}{d\eta}\left( \frac{a \mathcal{H}}{N} \right) \right)^{n-1}  \right) + .....
    \right. \nonumber\\
    &{} \left.
    + \left( d_0 - \sum \frac{n a}{N^{n+1}} d_n \left( \frac{\mathcal{H}}{\Lambda} \right)^n - \sum \frac{n}{ N^{n+1} \Lambda^{2n}} e_n \left( \frac{d}{d\eta}\left( \frac{a \mathcal{H}}{N} \right) \right)^n\right) \, (\partial_i A_j)^2  \right. \nonumber\\
    &{} \left. 
    - \frac{a^2 \mathcal{H}}{N^2} \frac{d}{d\eta} \left( (\partial_i A_j)^2 \sum \frac{n}{a N^n \Lambda^{2n}} e_n \left( \frac{d}{d\eta}\left( \frac{a \mathcal{H}}{N} \right) \right)^{n-1}  \right) + .....
    \right].
\end{align}

Setting $N(\eta) = a(\eta)$ in the above expression and using the truncated series expansion (\ref{eft:truncated}), we obtain the energy density for the EFT action:
\begin{align}
   \rho_{\rm EFT} = \rho_{\rm E} + \rho_{\rm B} + \rho_{\rm mixing}  
\end{align}
where 
\begin{subequations}
\label{eq:EFT-rho}
\begin{align}
    \rho_{\rm E} &= \frac{1}{a^4} \left( s_0 + \frac{s_1}{a} \frac{\mathcal{H}}{\Lambda} + \frac{1}{a^2} \left( \frac{\mathcal{H}}{\Lambda} \right)^2 
    \left[2 s_2 - b_1 (1 +  \epsilon_1) \right] \right) {A_i^{\prime}}^2 \\
    \rho_{\rm B} &= \frac{1}{a^4} \left( d_0 - \frac{d_1}{a} \frac{\mathcal{H}}{\Lambda} - 
    \frac{1}{a^2} \left( \frac{\mathcal{H}}{\Lambda} \right)^2 
    \left[2 d_2 - e_1 (1 + \epsilon_1) \right] \right) (\partial_i A_j)^2
    \\
    \rho_{\rm mixing} &= \frac{2}{a^6} \frac{\mathcal{H}}{\Lambda^2} \left(   b_1 A_i^{\prime} A_i^{\prime\prime} + e_1 \partial_i A_j \, \partial_i A_j^{\prime} \right)
\end{align}
\end{subequations}
where in the last expression $A_i^{\prime\prime}$ can be substituted by using the equations of motion (\ref{eq:GFieldEOM}) in the real space.

\iffalse
\section{EOM of $\delta A_{\mu}$}
\label{app:EOM}

Considering different scenarios the general equation of motion of $\delta A_{\mu}$ can be written as,

\begin{equation}
     \mathcal{A}_{i,k}^{\prime \prime} + \left(c_{\mathcal{A}}(t) k^2  + m^2(t) \right) \mathcal{A}_{i,k} = 0
\end{equation}

Here $\mathcal{A}_i$ is canonically normalised field and the normalisation factor will depend on the particular scenario we are considering. The sound speed $c_{\mathcal{A}}$ can appear from two effects: $1.$ from the coupling between scalar and $\mathcal{A}_i$ as discussed earlier, $2.$ the Gallileon scenario can also affect the sound speed.

Here, one important thing to note is that as $\pi A_i A_i$ coupling affects the sound speed, this put an automatic constraint on the mode function $\pi$. Because, $0< c_{\mathcal{A}}<1$, the mode functions $\pi$ has to obey $\pi < 1$.
\fi

\section{EFT magnetogenesis power-spectrum for power-law inflation}
\label{app:power-law}

In Sec. \eqref{sec:EFT-Spectrum}, we obtained the power-spectrum during slow-roll inflation. In this Appendix, we obtain the power spectrum for a generic power-law and de Sitter inflation. 

The procedure we follow to obtain the magnetic power-spectrum is the same as in the case of slow-roll. In the case of de Sitter, the power spectrum is exact while in the case of power-law inflation, we use WKB approximation. 
The scale factor during the power-law inflation is
\begin{equation}
    a(\eta) = (-H_0 \eta)^{\beta+1}.
\end{equation}
where $\beta \leq 2$ for power-law inflation, $\beta=-2$ corresponds to 
exact de Sitter and $H_0$ denotes the characteristic energy scale associated with inflation.

Like in Sec.~\eqref{sec:EFT-Spectrum}, 
to solve Eq. \eqref{eq:GFieldEOM}, we introduce new variables such that all the time dependence can be converted into momentum
dependence~\cite{2004-Shankaranarayanan.Sriramkumar-PRD}:
\begin{equation}\label{new-var}
x = \ln{\frac{\beta + 1}{k \eta}}, 
\quad \mathcal{A}_k = e^{-\frac{x}{2}} u_k,
\end{equation}
In terms of these new variables, $\mathcal{H}$ and its derivatives are given by:
\begin{eqnarray}\label{H-deriv}
 \mathcal{H} = \frac{a^{\p}}{a} = e^x k, & & \mathcal{H}^{\p} = \frac{a^{\p\p}}{a} - \left(\frac{a^{\p}}{a}\right)^2 = - \frac{1}{(\beta+1)} e^{2x}k^2,\\
\mathcal{H}^{\p\p} = \frac{2}{(\beta+1)^2} e^{3x} k^3, & & 
\mathcal{H}^{\p\p\p} = \frac{a^{\p\p\p\p}}{a}- 4 \frac{a^{\p} a^{\p\p\p}}{a^2}+12 \left(\frac{a^{\p}}{a}\right)^2 \frac{a^{\p\p}}{a} - 6 \left(\frac{a^{\p}}{a}\right)^4
= - \frac{6}{(\beta+1)^3} e^{4x} k^4 \nonumber.
\end{eqnarray}

In the redefined variables the sound speed \eqref{def:CA} and $Z''/Z$ become:
\begin{eqnarray}
\label{eq:cA-Powerlaw}
c_{A}^2 &=& 1+ (d_1-s_1) \frac{e^x}{a}  \frac{k}{\Lambda}+ 
\left[ (s_1^2-s_2-s_1 d_1+d_2) 
- \frac{(e_1-b_1)}{(\beta+1)} e^x \right]
\frac{e^x}{a^2}  
    \left(\frac{k}{\Lambda}\right)^2, 
    \\
\label{eq:Z-powerlaw}
\frac{Z^{\p\p}}{Z} &=& \frac{s_1}{2}\left\{1+\frac{3}{\beta+1}+\frac{2}{(\beta+1)^2}\right\} \frac{e^{3x}}{a}
    \frac{k^3}{\Lambda} \nonumber \\
& +& \left\{\frac{s_1^2}{4}+2s_2-\frac{s_1^2+5s_2-2b_1}{\beta+1}
    -\frac{\frac{s_1^2}{4}+2b_1}{(\beta+1)^2}-
    \frac{3b_1}{(\beta+1)^3}
    \right\} \frac{e^{4x}}{a^2}
    \frac{k^4}{\Lambda^2}.
\end{eqnarray}
where the scale factor $a(\eta)$ in 
the new variables is:
\begin{equation}\label{scale-factorx}
    a(x)=\left(H_0 e^{-x}\frac{\beta+1}{k}\right)^{\beta+1}.
\end{equation}
Substituting Eqs.~(\ref{eq:cA-Powerlaw}, \ref{eq:Z-powerlaw}) 
in Eq. \eqref{eq:GFieldEOM}, we have:
{\small
\begin{equation}\label{u-eom}
\frac{d^2 u_k}{d x^2} - 
\left[\frac{1}{4} 
- e^{-2x}(\beta+1)^2 
- \left\{M_1 e^{\beta x}+M_3 e^{(\beta+2)x}
+\left(M_2 
+M_4e^{2x}\right)e^{(2\beta+2)x}\frac{k^{\beta+2}}{\Lambda}\right\}
\frac{k^{\beta+2}}{\Lambda} \right]
u_k=0.
\end{equation}
}
where,
{\small
\begin{eqnarray}
M_1 &=& \frac{(\beta+1)^2}{\left(H_0 (\beta+1)\right)^{(\beta+1)}}(d_1-s_1),\\
M_2 &=& \frac{(\beta+1)^2}{\left(H_0 (\beta+1)\right)^{2(\beta+1)}}\left(s_1^2-s_2+s_1 d_1 +d_2-\frac{e_1-b_1}{\beta+1}\right),\\
M_3 &=& \frac{s_1}{2} \frac{(\beta+1)^2}{\left(H_0 (\beta+1)\right)^{(\beta+1)}} \left(1+\frac{3}{\beta+1}+\frac{2}{(\beta+1)^2}\right),\\
M_4 &=& \frac{(\beta+1)^2}{\left(H_0 (\beta+1)\right)^{2(\beta+1)}} 
      \left\{\frac{s_1^2+8 s_2}{4}-\frac{s_1^2-5s_2+2b_1}{\beta+1}
   -\frac{3s_1^2+4s_2+20 b_1}{4(\beta+1)^2}-
    \frac{3b_1}{(\beta+1)^3}\right\} \, .
\end{eqnarray}
}
Eq.~\eqref{u-eom} is the equation of motion for generic power-law inflation. 
It is not possible to obtain an exact analytical solution for a generic power law. Hence, in the rest of this appendix, we obtain the exact expression for the de Sitter case ($\beta = - 2$) and obtain the WKB solution for an arbitrary value of $\beta$.

\subsection{Power spectrum for de Sitter}

Setting $\beta=-2$ in \eqref{u-eom}, we have: 
\begin{equation}\label{eom-redefined}
\frac{d^2 u_k}{d x^2}+c_A^2 e^{-2x} u_k -m_2^2 u_k = 0 \, ,
\end{equation}
where,
\begin{eqnarray}
c_A^2 = 1+C_1 \frac{H_0}{\Lambda}+C_2 \frac{H_0^2}{\Lambda^2}\, , &~~~& 
m_2^2 = \frac{1}{4} -C_4 \frac{H_0^2}{\Lambda^2} \\
C_1 = s_1-d_1 \, , ~~C_4 = \frac{s_1^2}{2}-4 s_2 \, &~~~&
C_2 = s_1^2-s_2+s_1 d_1 +d_2 +e_1-b_1 \, .
\end{eqnarray}
As in the slow-roll case, the solution to Eq.~\eqref{eom-redefined} are Hankel functions~\cite{abramowitz+stegun}:
\begin{equation}
    u_k(x) = \alpha H_{m_2}^{(1)}(e^{-x}m_1) + \beta H_{m_2}^{(2)}(e^{-x}m_1),
\end{equation}
Following the steps from Eq.~\eqref{eq:EFT-Modefunction} 
to Eq.~\eqref{eq:BunchDaviesEFT}, 
the gauge-field mode functions ($A_k$) in the super-Hubble scales reduce to: 
\begin{equation}
A_k(\eta) = \sqrt{\frac{\pi}{4}}\left(\frac{-i}{\pi}\right) \Gamma(m_2) \, \frac{c_A}{Z}(-\eta)^{\frac{1}{2}}
\left(\frac{- c_A k\eta }{2}\right)^{-m_2},
\end{equation}
where,
\begin{equation}
   Z = 1 + d_1 \frac{H_0}{\Lambda}+(d_2+e_1) \frac{H_0^2}{\Lambda^2}
\end{equation}
The magnetic field power spectrum is:
\begin{equation}
\label{de-sitter-ps}
    \mathcal{P}_B = \frac{H^4}{8 \pi^3 Z^2 c_A^3} \Gamma^2(m_2) (-c_A k \eta)^{4+C_4\frac{H_0^2}{\Lambda^2}}
\end{equation}
This matches with the power-spectrum \eqref{PS-srFin} for slow-roll inflation by setting $\epsilon_1 = 0$.

\iffalse
\textcolor{red}{\textbf{Note}: Here one important thing to note is that in this power 
spectrum if both $C_3$ and $C_4$ are negligible with respect to the first term i,e 4 
in the exponent of $(c_A k \tau)$ then the power spectrum will be proportional to 
$\eta^4$ which corresponds to the fact that the power spectrum falls as
$\frac{1}{a^4}$ which is similar to the standard electromagnetic case. This is
actually consistent with the fact that in the EFT expansion the leading order term
i,e $s_0$ and $d_0$ are dominant over their EFT correction and ultimately the
Lagrangian can be approximated as the standard EM case unless we tune the parameters
$C_3$ or $C_4$ such that they are comparable to the leading order term.}
\fi

\subsection{Power spectrum for generic power-law inflation}

Unlike de Sitter, we can not obtain an exact analytical expression for generic power-law inflation. Instead, we use the WKB approximation to obtain the spectrum~\cite{2004-Shankaranarayanan.Sriramkumar-PRD}. Rewriting \eqref{u-eom} as
\begin{equation}
\frac{d^2 {u}_{k}}{dx^2}  + 
\omega^2(x)\, {u}_{k}(x) 
= 0,\label{eq:demukcux} 
\end{equation}
where $\omega^2(x)$ is 
\begin{equation}\label{gen-beta}
\omega^2(x)=e^{-2x}(\beta+1)^2+M_1 e^{\beta x}\frac{k^{\beta+2}}{\Lambda}
+M_2 e^{(2\beta+2)x}\frac{k^{2\beta+4}}{\Lambda^2}
+M_3 e^{(\beta+2)x}\frac{k^{\beta+2}}{\Lambda}
+M_4e^{(2\beta+4)x}\frac{k^{2\beta+4}}{\Lambda^2}-\frac{1}{4}    
\end{equation}
In the sub-horizon limit ($x\rightarrow -\infty$), the third term in RHS of the above expression dominates for $\beta<-2$. Hence, in the sub-horizon region ($I$), we have:
\begin{equation}
u^{\rm I}_{k}(x) = \frac{1}{\sqrt{\omega_I(x)}} \,
\left[ A_k \, \exp \, i\int^{x}dy 
\, \omega_I(y)
+ B_k \, \exp \, - i\int^{x}dy\, \omega_I(y) \right]
,\label{eq:wkb}
\end{equation}
where  ${\bar A}(k)$ and ${\bar B}(k)$ are $k$-dependent constants that
are to be fixed by the initial conditions, and
\begin{equation}
\omega_I^2(x) =  \frac{M_2}{\Lambda^2} \, e^{(2\beta+2)x} 
\, k^{2\beta+4} \, .
\end{equation}

As in the slow-roll and power-law cases, we shall assume that the gauge field is in the Bunch-Davies vacuum \eqref{eq:BunchDaviesEFT} on the sub-Hubble scales. This leads to:
\begin{eqnarray}
A_k =  -\frac{i (\beta +1)}{ \sqrt{32} \, \omega_I(x_i)} \, , 
&~~~&
B_k =\frac{1}{\sqrt{32} \, \omega_I(x_i)} \left(i \beta +4 \omega_I(x_i)+ i\right) \, . 
\end{eqnarray}
In the super-horizon limit ($x\rightarrow \infty$), the first term in the RHS of \eqref{gen-beta} dominates for $\beta < -2$. Hence, 
the super-horizon region (II), we have:
\begin{equation}\label{region-IIWKB}
    u^{\rm II}_k(x) = \frac{C_k}{\sqrt{\omega_{\rm II}(x)}} \exp{\left( \int_x^{x_f} \vert\omega_{\rm II}(y)\vert dy\right)} + \frac{D_k}{\sqrt{\omega_{\rm II}(x)}} \exp{\left( \int_x^{x_f} -\vert\omega_{\rm II}(y)\vert dy\right)}
\end{equation}
Near the horizon crossing ($x=0$), the WKB approximation breaks down and in this region, we can approximate $\omega(x) = \alpha \, x$, 
where $\alpha$ depends on $M_1, M_2, M_3, \Lambda$ and $\beta$. 
Matching the modes at the horizon exit, we have the following relation between $C_k, D_k$ with $A_k, B_k$:
\begin{eqnarray}
C_k = \frac{A_k+B_k}{\sqrt{8}} e^{-\Psi};~~D_k = \frac{B_k-A_k}{\sqrt{2} i} e^{\Psi} \,
\end{eqnarray}
where,
\begin{equation}
    \Psi = \int_{x_*}^{x_f}\alpha \, y \, dy
\end{equation}
Taking into account only the growing mode, the modulus squared of the mode function is:
\begin{equation}\label{powerlaw-spec}
\left\vert\mathcal{A}_k\right\vert^2 = \frac{\left\vert B_k-A_k\right\vert^2}{2} e^{-x}\frac{e^{2\Psi}}{\sqrt{\omega(x)}}.
\end{equation}
Substituting this in Eq.~\eqref{eq:pMS}, the magnetic power spectrum at the super-horizon scales is:
\begin{equation}
    \mathcal{P}_B (k) = \frac{k^5}{2 \pi^2 a^4} \frac{\left\vert B_k-A_k\right\vert^2}{2 Z} e^{-x}\frac{e^{2\Psi}}{\sqrt{\omega(x)}}
\end{equation}
%
%
%============ Vector Galileon Appendix =========
%
\section{Vector Galileon: Equation of motion in terms of Slow-roll parameters}
\label{app:VG-slow-roll}

In case of slow-roll inflation, the slow-roll parameters are defined
as~\cite{2014-Martin.etal-PhyDarkUniv}:
\begin{eqnarray}
\epsilon_1 = -\frac{H^{\prime}}{ a H^2} = 1 - \frac{\mathcal{H}^{\prime}}{\mathcal{H}^2};~~~
&&\epsilon_{2} = \frac{\epsilon^{\prime}_1}{a H \epsilon_1} 
= \frac{ \left( 2{\mathcal{H}^{\prime}}^2 - \mathcal{H}\mathcal{H}^{\prime\prime} \right) }{\mathcal{H}^2 \left(\mathcal{H}^2 - \mathcal{H}^{\prime} \right) } ~~.
\label{def:Slowrollpara}
\end{eqnarray}
%
%which gives, 
%\begin{eqnarray}
%a^{\prime} = a^2 H \\
%a^{\prime\prime} = - \frac{(\epsilon_1 - 2) a^{\prime}{}^2 }{a} \\
%\end{eqnarray}
%
The equation of motion (\ref{eq:eom-canonicalA}) for the canonical vector field $\mathbb{A}_k$  in terms of slow-roll parameters can be obtained as
\begin{align}\label{eqapp:eom-canonical-SR}
\mathbb{A}_k^{\prime\prime} +  \left[ c_s^2 \, k^2 -    \frac{ {\mathcal{H}}^2 J \epsilon_1}{(1 - J )^2} \, \left(1 - J - 3\epsilon_1 + 2 J\epsilon_1 + \epsilon_2 (1 - J)  \right) \right] \,\mathbb{A}_k = 0 \, ,
\end{align}
where $c_s$ is defined in Eq. \eqref{eq:def-cs}.
Substituting Eq.(\ref{eq:def-canonicalA}) in Eq.~(\ref{eqapp:eom-canonical-SR}), the equation of motion of the physical vector field $A_{k}$ is:
%for the action (\ref{eq:eom-canonicalA-fourier})
%
\begin{align}
A_k^{\prime\prime} + \frac{2 \mathcal{H} J \epsilon_1 }{ (1 - J) } A_k^{\prime} 
+ \frac{ k^2 (1 - J + J \epsilon_1 ) }{ (1 - J) } A_k = 0 \, .
\end{align}

\acknowledgments
The authors thank Avijit Chowdhury, Joseph P Johnson, Karthik Rajeev and Urjit Yajnik for the comments on the earlier draft.  AK is supported by the MHRD fellowship at IIT Bombay. This work is supported by ISRO Respond grant. DN is supported by INSPIRE-Faculty fellowship of DST-SERB, India. Finally, we would like to thank the referee for his or her insightful comments and suggestions that have helped to improve the manuscript.

%\paragraph{Note added.} This is also a good position for notes added after the paper has been written.

\bibliographystyle{unsrt}
%\bibliography{References}

\end{document}